\documentclass[a4paper,12pt]{article}

\usepackage[english]{babel}
\usepackage{graphicx}
\usepackage{colordvi}
\usepackage{fancybox}
\usepackage{cancel}
\usepackage{amsmath,amsfonts}
\usepackage{comment}
\usepackage{longtable}
\usepackage{booktabs}
\usepackage{hyperref}

\usepackage{color}
\definecolor{gray}{rgb}{0.5,0.5,0.5}

\newcommand\lsim{\mathrel{\rlap{\lower4pt\hbox{\hskip1pt$\sim$}}
    \raise1pt\hbox{$<$}}}
\newcommand\gsim{\mathrel{\rlap{\lower4pt\hbox{\hskip1pt$\sim$}}
    \raise1pt\hbox{$>$}}}

\newcommand{\beq}{\begin{equation}}
\newcommand{\eeq}{\end{equation}}
\newcommand{\bea}{\begin{eqnarray}}
\newcommand{\eea}{\end{eqnarray}}
\newcommand{\bem}{\begin{pmatrix}}
\newcommand{\eem}{\end{pmatrix}}
\newcommand{\noi}{\noindent}
\def\ydt{Y_{\widetilde{27}}}

\def\yd{Y_{27}}
\def\yt{Y_{\overline{351}'}}
\def\SU{\mathrm{SU}}
\def\SO{\mathrm{SO}}
\def\RED{\textcolor{red}}
\def\BLUE{\textcolor{blue}}

\begin{document}

\begin{flushright}
CETUP2013-019
\end{flushright}

\bigskip

\begin{center}

{\Large\bf  Towards the minimal renormalizable supersymmetric $E_6$ model}

\vspace{1cm}

\centerline{
Borut Bajc$^{a,b,}$\footnote{borut.bajc@ijs.si} and
Vasja Susi\v c$^{a,}$\footnote{vasja.susic@ijs.si}
}
\vspace{0.5cm}
\centerline{$^{a}$ {\it\small J.\ Stefan Institute, 1000 Ljubljana, Slovenia}}
\centerline{$^{b}$ {\it\small Department of Physics, University of Ljubljana, 1000 Ljubljana,
Slovenia}}

\end{center}

\bigskip

\begin{abstract}

We find an explicit renormalizable supersymmetric $E_6$ model with all the ingredients
for being realistic. It consists of the Higgs sector $351'+\overline{351'}+27+\overline{27}$,
which breaks $E_6$ directly to the Standard Model gauge group.
Three copies of $27$ dimensional representations then describe the matter sector,
while an extra $27+\overline{27}$ pair is needed to successfully split the Standard Model
Higgs doublet from the heavy Higgs triplet. We perform the analysis of the vacuum structure and the Yukawa sector of this model, as well
as compute contributions to proton decay. Also, we show why some other simpler $E_6$ models fail to be realistic at the renormalizable level.

\end{abstract}

\clearpage

\tableofcontents



\section{Introduction}

In spite of the $E_6$ group being introduced \cite{Gursey:1975ki} quite soon after the first attempt of grand unification
\cite{Georgi:1974sy}, to our knowledge there exists no complete model in the literature so far.
With complete we mean not only the Yukawa sector, which has been studied occasionally
\cite{Gursey:1978fu,Achiman:1978vg,Shafi:1978gg,Ruegg:1979fr,Kalashnikov:1979dq,Barbieri:1980vc,Shaw:1980ku,Stech:2003sb,Caravaglios:2005gf,Caravaglios:2006aq,Stech:2008wd}, but also
the whole Higgs sector and symmetry breaking (exceptions being the simplest renormalizable \cite{Buccella:1987kc}
or non-renormalizable \cite{Bertolini:2010yz} supersymmetric case or the renormalizable non-supersymmetric
example in \cite{Fradkin:1977mj,Kalashnikov:1979dq}), as well as the
determination of the Standard Model (SM) Higgs doublet, i.e. the doublet-triplet splitting. This state of affairs is
probably due to two reasons. The first one is the complexity of the $E_6$ group, which is a bit less familiar than
the $\SU(N)$ \cite{Georgi:1974sy} or $\SO(N)$ \cite{Wilczek:1981iz} structures that can be easily generalized
from simpler low
dimensional cases. The second one is the lack of a serious motivation. The fact that the quantum numbers
of the fundamental $27$-dimensional representation can accommodate both the matter $16$ and the Higgs
$10$ of $\SO(10)$ is nice in principle, but not easy to make it useful and realistic. Not only that, but at least part of
what one gains coming from $\SU(5)$ to $\SO(10)$, for example automatic R-parity conservation in $\SO(10)$
\cite{Mohapatra:1986su,Aulakh:1999cd,Aulakh:2000sn} with
$126$ breaking rank \cite{Aulakh:1982sw,Clark:1982ai,Aulakh:2003kg}, is lost when the same $\SO(10)$ is
embedded in $E_6$. Also, the simplest supersymmetric
model with
the lowest dimensional Higgs representations $27+\overline{27}+78$ can break at the renormalizable level
only to $\SO(10)$ \cite{Buccella:1987kc}. It is the purpose of this paper to fill this gap and present a fully realistic $E_6$ grand
unified theory (GUT). In doing this we simplify our analysis by assuming renormalizability and supersymmetry.
Contrary to the approach in \cite{Bajc:2008vk,Bajc:2012sm},
we will not consider the orthogonal problem of supersymmetry breaking in our model.

The simplest realistic model we were able to find is made out of  $351'+\overline{351'}+27+\overline{27}$,
needed to break $E_6\to$SU(3)$\times$SU(2)$\times$U(1), an extra $27+\overline{27}$ to achieve the
doublet-triplet splitting and so determine explicitly where the SM Higgs doublets live, and three copies of
the fundamental $27$ for the matter sector. Although we will not present a complete proof that this is
really the minimal or simplest renormalizable version, we will give various arguments for why some other
representations like $78$, $351$ or extra $27$'s in many cases cannot do the same job.

This paper is an extended and detailed version of \cite{Bajc:2013nta}.
We will start first in section~\ref{section:group-theory} with some general description of the $E_6$ representations, maximal
subgroups, and how to construct invariants. In section~\ref{section:examples} we will show why some examples cannot work,
while in section~\ref{section:realistic-Higgs-sector} we will present the minimal Higgs sector, and show the explicit solution which spontaneously
breaks the gauge group to the SM one. Section~\ref{section:MSSM-Higgses} will be devoted to the doublet-triplet splitting and the need
for an extra fundamental-antifundamental pair. Section~\ref{section:Yukawa-sector} will be devoted to the Yukawa sector, and its
peculiarities: the matter fields consist in $\SU(5)$ language of 3 generations of $10+\bar 5$, 3 vector-like
$5+\bar 5$ pairs, which, after being integrated out, lead to the needed flavor structure, and 3 pairs of $\SU(5)$ singlets.
Section~\ref{section:proton-decay} is an analysis of the contributions to $D=5$ proton decay in our model.
We will conclude in section~\ref{section:conclusions} with a list of open problems and possible future projects. Three appendices will give computational
details on the issues of state identification, checks on the vacuum solution and the doublet-triplet splitting.

Note on convention: for greater clarity, we color code the vacuum expectation values according to their mass scale: \RED{red} signifies a GUT mass scale, while \BLUE{blue} signifies an Electroweak scale.

\section{All we need to know about $E_6$\label{All-e6}\label{section:group-theory}}

$E_6$, similarly to the $\SU(N)$ groups, has two type of tensor indices: the fundamental or upper index, and
the anti-fundamental or lower index. They both run from $1$ to $27$, which is the dimensionality of the
fundamental and anti-fundamental representations. Tensors are constructed with these indices, and extra
constraints like simmetricity or antisymmetricity can be further imposed to get irreducible representations. Finally,
similarly to the case of the completely antisymmetric SU(N) invariant Levi-Civita tensor $\epsilon_{\alpha_1\ldots\alpha_N}$ or $\epsilon^{\alpha_1\ldots\alpha_N}$, we have in $E_6$ the 3-index completely symmetric
invariant tensors $d_{\mu\nu\lambda}$ and $d^{\mu\nu\lambda}$ with $\mu,\nu,\lambda=1\ldots27$.

The lowest dimensional ($<1000$) nontrivial representations are \cite{Slansky:1981yr}

\bea
27^\mu&\ldots&{\rm fundamental}\\
\overline{27}_\mu&\ldots&\textrm{anti-fundamental}\\
{78^\mu}_\nu&\ldots&{\rm adjoint}\;(={(t^A)^\mu}_\nu78^A)\\
351^{\mu\nu}=-351^{\nu\mu}&\ldots&{\rm two\;indices\;antisymmetric}\\
\overline{351}_{\mu\nu}=-\overline{351}_{\nu\mu}&\ldots&{\rm two\;indices\;antisymmetric}\\
351'^{\mu\nu}=+351'^{\nu\mu}&\ldots&{\rm two\;indices\;symmetric}\; (d_{\lambda\mu\nu}351'^{\mu\nu}=0)\\
\overline{351'}_{\mu\nu}=+\overline{351'}_{\nu\mu}&\ldots&{\rm two\;indices\;symmetric}\; (d^{\lambda\mu\nu}\overline{351'}_{\mu\nu}=0)\\
{650^\mu}_\nu&\ldots& ({650^\mu}_\mu={(t^A)^\nu}_\mu{650^\mu}_\nu=0)
\eea

\noi
with $t^A$ the generators of the $E_6$ algebra, with the adjoint indices $A=1\ldots78$ and
the fundamental indices $\mu,\nu=1\ldots27$. Note that our convention for labeling representations exchanges $351$ and $\overline{351}$, as well as $351'$ and $\overline{351'}$ compared to \cite{Slansky:1981yr}; in our convention, the representations without bars contain fundamental (upper) indices, while the barred representations contain antifundamental (lower) indices.

To get an explicit form for the invariant d-tensors, we can follow \cite{Kephart:1981gf}. We organize
the fields in $27$ according to their quantum numbers: we introduce three $3\times 3$ matrices $L$,
$M$, $N$, which contain all the fields in $27$ and which under the $\SU(3)_C\times\SU(3)_L\times\SU(3)_R$ maximal subgroup of $E_6$ transform as

\begin{align}
L&\sim(3,3,1),\\
M&\sim(1,\bar 3,3),\\
N&\sim(\bar 3,1,\bar 3).
\end{align}
\noindent
Then
\begin{align}
\tfrac{1}{6} d_{\mu\nu\lambda}27^\mu27^\nu27^\lambda\equiv -\det{L}+\det{M}-\det{N}-\mathrm{tr}(LMN).\label{d-tensor-equation}
\end{align}

Note that the first and third terms on the right have a minus sign compared to \cite{Kephart:1981gf} due to the different embedding of the $\SU(3)_L$ and $\SU(3)_R$ parts of the maximal subgroup $\SU(3)_C\times\SU(3)_L\times\SU(3)_R$. Our embedding conforms to the one in \cite{Slansky:1981yr}, which is more useful from the physics point of view. Another difference from \cite{Kephart:1981gf} is the factor $\tfrac{1}{6}=\tfrac{1}{3!}$ in front of $d_{\mu\nu\lambda}$ on the left, which is needed to ensure the normalization

\begin{align}
d_{\mu\lambda\rho} d^{\lambda\rho\nu}&=10 \delta_{\mu}{}^{\nu}
\end{align}

\noi
claimed in \cite{Kephart:1981gf}. The tensor $d^{\mu\nu\lambda}$ with all upper indices is taken to have the same numerical values as the tensor $d_{\mu\nu\lambda}$ with lower indices.

With the above definition and normalization, the only nonzero values of the $d$-tensor are either $1$ or $-1$. Another important property of the $d$-tensor can be deduced from equation~\eqref{d-tensor-equation}: although $d$ is symmetric in its indices, it gives zero as soon as two of the three indices take the same value, similar to the completely antisymmetric tensors $\varepsilon_{\alpha_1\ldots\alpha_N}$.

With all this we can now see some explicit examples of models.

\section{Some simple unsuccessful examples\label{section:examples}}

We assign the SM matter fields to three copies of the $27$-dimensional representation. In order to avoid possible
issues with R-parity  breaking nonzero vacuum expectation values in the $16$ of $\SO(10)$, we will enforce a $\mathbb{Z}_2$ parity, under which the
matter $27$'s are odd, while all other "Higgs" representations are even. In this and in the next section, we will consider various Higgs sectors and assess, whether they enable a direct breaking of $E_6$ to the SM group $\SU(3)_C\times\SU(2)_L\times \mathrm{U}(1)_Y$.

\subsection{$n_{27}\times27+n_{\overline{27}}\times\overline{27}+n_{78}\times 78$\label{section-272778}}

Trying to build realistic renormalizable models just from the fundamental, antifundamental and adjoint representations $27$, $\overline{27}$ and $78$, possibly in multiple copies, proves to be impossible due to group-theoretic reasons alone. We shall describe these reasons below where we label the number of copies of the representations $27$, $\overline{27}$ and $78$ in the model by $n_{27}$, $n_{\overline{27}}$ and $n_{78}$ respectively.

In order to break $E_6$ to the SM gauge group, only SM singlets can acquire a nonzero vacuum expectation value
(VEV). First note the $\SO(10)$ decompositions of the representations under consideration:

\begin{align}
27&=16+10+1,\\
\overline{27}&=\overline{16}+10+1,\\
78&=45+16+\overline{16}+1.
\end{align}

Only the representations $1$, $16$, $\overline{16}$ and $45$ of $\SO(10)$ contain SM singlets. The SM singlets of $1$, $16$ and $\overline{16}$ are also $\SU(5)$ singlets, while the $45$ contains both a $1$ and a $24$ of $\SU(5)$. In total, the representation $27$ contains $2$ SM singlets, both of which are also $\SU(5)$ singlets, while the $78$ contains $5$ singlets, $4$ of which are $\SU(5)$ singlets and one is in a $24$ of $\SU(5)$.

Taking $n_{78}=0$, we are left only with $\SU(5)$ singlets in the $27$'s and $\overline{27}$'s, so $\SU(5)$ remains unbroken, regardless of the number of $27$ and $\overline{27}$ copies. To break the $\SU(5)$ part to the SM, we need a model with at least one $78$, which acquires a nonzero VEV in the $24$ of $\SU(5)$. We will show below that no $\langle 24\rangle$ in a $78$ can be nonzero, no matter how many $27$, $\overline{27}$ and $78$ copies in the renormalizable model, and thus $\SU(5)$ remains unbroken.

Consider first all types of invariants, which can be formed from the representations $27_i$, $\overline{27}_j$ and $78_k$. Besides the mass terms, we have the following cubic invariants:
\begin{align}
27_i\times 27_j\times 27_k,\\
\overline{27}_i\times \overline{27}_j\times \overline{27}_k,\\
27_i\times 78_k\times\overline{27}_j,\label{model1-invariant3}\\
78_i\times 78_j\times 78_k.\label{model1-invariant4}
\end{align}

The only relevant invariants are the last two, since they alone contain a $\langle 24\rangle\subset 78$. We analyze which combinations of VEVs can form terms of these invariants.

The invariant~\eqref{model1-invariant3} does not contain a term with the $\langle 24\rangle$ of $78$, since the $27$ contains only $\SU(5)$ singlets and $1\times 24\times 1$ does not contain a singlet in the $\SU(5)$ language, so this term cannot be present in the invariant.

The cubic invariant~\eqref{model1-invariant4} is antisymmetric in the $78$-factors, so we need at least three different $78$ copies in the model in order for this invariant to be nonzero. Assuming $n_{78}\geq 3$, the $24$'s can enter into the invariant in
two possible factor combinations in the $\SU(5)$ language: $1\times 24\times 24$ and $24\times 24\times 24$. But only symmetric products of the $24$ form an $\SU(5)$ invariant, while our case is antisymmetric due to the cubic $78^3$ being antisymmetric. Again, no terms containing any $\langle24\rangle$'s are present in the invariant.

Since no cubic invariants contain the $\langle 24\rangle$'s, these VEVs only have their mass terms, which force us to $\langle 24\rangle=0$. Only $\SU(5)$ singlets can therefore acquire VEVs and $\SU(5)$ remains unbroken. Note that the $\langle 24\rangle$ as a part of $78$ in $E_6$ behaves differently from the $\langle 24\rangle$ in $\SU(5)$. In $\SU(5)$, the $\langle 24\rangle$ can acquire a VEV due to the existence of both the quadratic and cubic invariants $\mathrm{Tr}\,24^2$ and $\mathrm{Tr}\,24^3$. In $E_6$, it is the cubic term which is missing due to the antisymmetry of $78^3$: $\mathrm{Tr}\, 78^3=0$.

This conclusion also applies to the special case $n_{27}=n_{\overline{27}}=n_{78}=1$ in the literature~\cite{Buccella:1987kc}. This renormalizable model breaks $E_6$ to $\SO(10)$, which contains the invariant $\SU(5)$. Our result shows that adding extra copies of representations to this model, while still at the renormalizable level, will never allow breaking to the SM.


\subsection{$351+\overline{351}+n_1\times 27+n_2\times\overline{27}$}

The representation $351$ is a two index antisymmetric representation.
It decomposes under $\SO(10)$ as
\begin{align}
351&=10+16+\overline{16}+45+120+\overline{144}.
\end{align}
From this we conclude that the $351$ contains $5$ SM singlets, $3$ of which are $\SU(5)$ singlets (in $16$, $\overline{16}$ and $45$ of $\SO(10)$) and $2$ are part of $24$ under $\SU(5)$ (in $45$ and $\overline{144}$ of $\SO(10)$).

Although $351$ forms a cubic invariant $351^3$, it is antisymmetric in the $351$ factors. In the simplest models with only one copy of the pair $351+\overline{351}$, the cubic invariants are trivially zero.
The renormalizable superpotential of the model $351+\overline{351}$ therefore contains only the mass term $351\times \overline{351}$. The $F$-terms then give all VEVs to be zero and no breaking occurs.

Adding pairs of $27+\overline{27}$, we have the presence of invariants

\begin{align}
27^\mu\; 27^\nu\; \overline{351}_{\mu\nu},\label{invariant-351b2727}\\
\overline{27}_\mu\; \overline{27}_\nu\; 351^{\mu\nu}.\label{invariant-35127b27b}
\end{align}

Note that due to the antisymmetry of $351$ and $\overline{351}$, these invariants are trivially zero if we have just a single copy of $27$ and $\overline{27}$.

Assume we have more than a single copy of $27$ and $\overline{27}$ so that the invariants~\eqref{invariant-351b2727} and \eqref{invariant-35127b27b} are nonzero. Since all VEVs in the $27$ are $\SU(5)$ singlets, it is up to the two $24$'s in $351$ (and two in the $\overline{351}$) to break $\SU(5)$. But similarly to the models in section~\ref{section-272778}, the $24$'s of $351$ and $\overline{351}$ are again present only in the mass cross-term: the cubic invariants $351^3$ and $\overline{351}^3$ are trivially zero, while the invariants~\eqref{invariant-351b2727} and \eqref{invariant-35127b27b} do not contain the $24$'s, since the only term with a $24$ in the invariant, written in $\SU(5)$ parts containing VEVs, could be $1\times 1\times 24$, which is not invariant under $\SU(5)$.

In models with $351+\overline{351}$ and an arbitrary number of $27$'s and $\overline{27}$'s, the $24$'s in the $351+\overline{351}$ never acquire VEVs and consequently $\SU(5)$ remains unbroken.

\subsection{$351'+\overline{351'}$\label{section:351'-model}}

The representation $351'$ is a two index symmetric representation, with a cubic invariant $351'^3$ symmetric in its factors.

In contrast with the previous models, we cannot discount this model by simple group-theoretic arguments alone. An explicit computation shows this model breaks $E_6$ to the Pati-Salam (PS) group $\SU(2)_L\times\SU(2)_R\times\SU(4)_C$. Since this result is a special case of our proposed model, we postpone the discussion of this model until section~\ref{section:alternative-solutions}.

\section{A realistic Higgs sector\label{section:realistic-Higgs-sector}}
\subsection{The model: $351'+\overline{351'}+27+\overline{27}$}
We find that the combination $351'+\overline{351'}+27+\overline{27}$ forms a realistic Higgs sector, which breaks $E_6$ to the SM.

First, note the decomposition of the representation $351'$ under $\SO(10)$:

\begin{align}
351'&=1+10+16+54+126+\overline{144}.
\end{align}
This representation contains $5$ SM singlets, $3$ of which are $\SU(5)$ singlets (in $1$, $16$ and $126$ of $\SO(10)$), and $2$ are part of a $24$ under $\SU(5)$ (in $54$ and $\overline{144}$ of $\SO(10)$). The Higgs sector $351'+\overline{351'}+27+\overline{27}$ therefore contains $5+5+2+2=14$ singlets in total. We list them in Table~\ref{table-singlets}.

\begin{table}[h]
\caption{SM singlet VEVs in our Higgs sector.\label{table-singlets}}
\vskip 0.2cm
\centering
\scalebox{0.9}{
\begin{tabular}{rrrrr@{\hspace{1.5cm}}rrrrr}
\toprule
label&
$\subseteq\mathrm{PS}$&
$\subseteq\SU(5)$&
$\subseteq\SO(10)$&
$\subseteq \mathrm{E}_6$&
label&
$\subseteq\mathrm{PS}$&
$\subseteq\SU(5)$&
$\subseteq\SO(10)$&
$\subseteq \mathrm{E}_6$\\\midrule
$c_1$&$(1,1,1)$&      $1$    &$1$                &$27$&$d_1$&$(1,1,1)$&$1$    &$1$                &$\overline{27}$\\
$c_2$&$(1,2,\overline{4})$&      $1$    &$16$               &$27$&$d_2$&$(1,2,4)$&$1$    &$\overline{16}$    &$\overline{27}$\\\addlinespace
$e_1$&$(1,3,10)$&$1$&$126$&$351'$&$f_1$&$(1,3,\overline{10})$&$1$&$\overline{126}$&$\overline{351'}$\\
$e_2$&$(1,2,\overline{4})$&$1$&$16$&$351'$&$f_2$&$(1,2,4)$&$1$&$\overline{16}$&$\overline{351'}$\\
$e_3$&$(1,1,1)$&$1$&$1$&$351'$&$f_3$&$(1,1,1)$&$1$&$1$&$\overline{351'}$\\
$e_4$&$(1,1,1)$&$24$&$54$&$351'$&$f_4$&$(1,1,1)$&$24$&$54$&$\overline{351'}$\\
$e_5$&$(1,2,4)$&$24$&$\overline{144}$&$351'$&$f_5$&$(1,2,\overline{4})$&$24$&$144$&$\overline{351'}$\\\bottomrule
\end{tabular}
}
\end{table}

Note that the $c$'s and $d$'s denote the VEVs in $27$ and $\overline{27}$, as in \cite{Buccella:1987kc}, while the $e$'s and $f$'s denote the VEVs in $351'$ and $\overline{351'}$, respectively. The VEVs $\langle 24\rangle$ under $\SU(5)$ are $e_4$, $e_5$, $f_4$ and $f_5$. All singlets have the standard  K\"{a}hler normalization

\begin{align}
\langle 27^\mu\;27^*_\mu\rangle&=|c_1|^2+|c_2|^2,\\
\langle\overline{27}_\mu\;\overline{27}^{\ast\mu} \rangle&=|d_1|^2+|d_2|^2,\\
\langle 351'^{\mu\nu}\,351^{\prime \ast}_{\mu\nu} \rangle&=|e_1|^2+|e_2|^2+|e_3|^2+|e_4|^2+|e_5|^2,\\
\langle\overline{351'}_{\mu\nu}\,\overline{351'}^{*\mu\nu} \rangle&=|f_1|^2+|f_2|^2+|f_3|^2+|f_4|^2+|f_5|^2.
\end{align}

The most general renormalizable superpotential of the model can be written as
\begin{align}
W&=m_{351'}\;I_{351'\times\overline{351'}}+m_{27}\;I_{27\times\overline{27}}\nonumber\\
&\qquad + \lambda_1 \;I_{351'^3}+\lambda_2 \;I_{\overline{351'}^3}+\lambda_3 \;I_{27^2\times \overline{351'}}+\lambda_4 \;I_{\overline{27}^2\times 351'} +\lambda_5 \;I_{27^3} + \lambda_6 \;I_{\overline{27}^3}.\label{equation-superpotential}
\end{align}

Explicit computation yields the following expressions for the superpotential invariants:
        \begin{align}
        I_{351'\times\overline{351'}}&=\overline{351'}_{\mu\nu}\;351^{\mu\nu}=\RED{e_1} \RED{f_1}+\RED{e_2} \RED{f_2}+\RED{e_3} \RED{f_3}+\RED{e_4} \RED{f_4}+\RED{e_5} \RED{f_5},\\
        I_{27\times\overline{27}}&=\overline{27}_\mu\; 27^\mu=\RED{c_1} \RED{d_1}+\RED{c_2} \RED{d_2},\\
        I_{351'^3}&=351'^{\mu\alpha}\;351'^{\nu\beta}\;351'^{\lambda\gamma}\;d_{\alpha\beta\gamma}d_{\mu\nu\lambda}=3 \left(\RED{e_3} \RED{e_4}^2+\RED{e_1} \RED{e_5}^2 - \sqrt{2} \RED{e_2} \RED{e_4} \RED{e_5} \right),\\
        I_{\overline{351'}^3}&=\overline{351'}_{\mu\alpha}\;\overline{351'}_{\nu\beta}\;\overline{351'}_{\lambda\gamma}\;d^{\alpha\beta\gamma}\;d^{\mu\nu\lambda}=3 \left(\RED{f_3} \RED{f_4}^2+\RED{f_1} \RED{f_5}^2 - \sqrt{2}\RED{f_2} \RED{f_4} \RED{f_5}\right),\\
        I_{27^2\times\overline{351'}}&=\overline{351'}_{\mu\nu}\;27^\mu\;27^\nu=\RED{c_2}^2 \RED{f_1} + \sqrt{2} \RED{c_1} \RED{c_2} \RED{f_2} + \RED{c_1}^2 \RED{f_3},\\
        I_{\overline{27}^2\times 351'}&=351'^{\mu\nu}\;\overline{27}_\mu\;\overline{27}_\nu=\RED{d_2}^2 \RED{e_1} + \sqrt{2} \RED{d_1} \RED{d_2} \RED{e_2} + \RED{d_1}^2 \RED{e_3},\\
        I_{27^3}&=27^\mu\;27^\nu\;27^\lambda\;d_{\mu\nu\lambda}=0,\\
        I_{\overline{27}^3}&=\overline{27}_\mu\;\overline{27}_\nu\;\overline{27}_\lambda\;d^{\mu\nu\lambda}=0.\label{equation-invariant-end}
        \end{align}

Note that
 the invariants $27^3$ and $\overline{27}^3$ are not trivially zero, they just do not contain any terms with only SM singlets, which are relevant for the equations of motion. The zero result can be easily understood under the decomposition $\SO(10)\times\mathrm{U}(1)''\subset E_6$: by noting that the $\mathrm{U}(1)''$ quantum numbers of both singlets in $27$ have the same sign, no triple product of the two VEV singlets will yield a net $\mathrm{U}(1)''$ charge to be zero, which is a requirement for the term to be present in the invariant. Alternatively, from the point of view of the invariant $d$-tensor, since there are only $2$ SM singlets in the $27$, only components with two same-value indices of the $d$-tensor contribute, but these are zero, as already mentioned in section~\ref{All-e6}. Analogous considerations apply to the invariant $\overline{27}^3$.

\subsection{Equations of motion}

Our model is supersymmetric, so the equation of motion consist of both $F$-terms and $D$-terms. The $F$-terms are determined by the equations
\begin{align}
0&=\frac{\partial W}{\partial c_i}=\frac{\partial W}{\partial d_i}=\frac{\partial W}{\partial e_j}=\frac{\partial W}{\partial f_j},
\end{align}
with $i=1,2$ and $j=1,\ldots,5$. The $F$ terms give in total $14$ holomorphic equations containing $14$ holomorphic variables. Since we have already written the superpotential $W$ explicitly in equations~\eqref{equation-superpotential}-\eqref{equation-invariant-end}, the equations of motion can be trivially derived by the reader.

The $D$-terms on the other hand, take the form

\begin{align}
D^A&= \phantom{+}(27^\dagger)_\mu\;(\hat{t}^A\, 27)^\mu+(\overline{27}^\dagger)^\mu\;(\hat{t}^A\, \overline{27})_\mu\nonumber\\
&\quad+(351'^\dagger)_{\mu\nu}\;(\hat{t}^A\, 351')^{\mu\nu}+(\overline{351'}^\dagger)^{\mu\nu}\;(\hat{t}^A\, \overline{351'})_{\mu\nu},
\end{align}

\noindent
with the representations $27$, $\overline{27}$, $351'$ and $\overline{351'}$ containing the $14$ SM singlet VEVs. The index $A=1,\ldots,78$ is the $E_6$ adjoint index and the $\hat{t}^A$ is the action of the $A$-th algebra generator on the states in a given representation. As per the usual tensor methods in group theory, the actions of the $A$-th generator on different representations is

\begin{align}
(\hat{t}^A\, 27)^\mu&=\phantom{-}(t^A)^{\mu}{}_{\lambda}\; 27^\lambda,\\
(\hat{t}^A\, \overline{27})_\mu&=-(t^{A\ast})_{\mu}{}^{\lambda}\;\overline{27}_\lambda,\\
(\hat{t}^A\, 351')^{\mu\nu}&= \phantom{-}(t^A)^\mu{}_{\lambda}\; 351'^{\lambda\nu}+(t^A)^{\nu}{}_{\lambda}\; 351'^{\mu\lambda},\\
(\hat{t}^A\, \overline{351'})_{\mu\nu}&=-(t^{A\ast})_\mu{}^{\lambda}\; \overline{351'}_{\lambda\nu} -(t^{A\ast})_\nu{}^{\lambda}\; \overline{351'}_{\mu\lambda},
\end{align}
where the symbol $(t^A)^{\mu}{}_{\nu}$ denotes the components of the $A$-th generator as a $27\times 27$ matrix, and ${}^\ast$ denotes complex conjugation.

Out of possible $78$ $D$-terms, only $5$ are non-trivial and they correspond exactly to the $5$ SM singlets in the $78$. The singlet generators are the following generators of the maximal subgroup $\SU(3)_C\times\SU(3)_L\times\SU(3)_R$ of $E_6$: $t_L^8$, $t_R^3$, $t_R^6$, $t_R^7$ and $t_R^8$. We therefore label the nonzero $D$-terms accordingly: $D_L^8$, $D_R^3$, $D_R^6$, $D_R^7$ and $D_R^8$.  Furthermore, the combination $D_L^8+\sqrt{3}D_R^3+D_R^8$ of the non-trivial $D$-terms is trivially zero, since this combinations of generators corresponds to the SM hypercharge generator $t^Y$:
\begin{align}
t^Y&=t_L^8+\sqrt{3} t_R^3+t_R^8.
\end{align}
The independent equations for the $D$-terms can be further simplified by taking their linear combinations. The $4$ independent $D$-term real constraints can be written as
\begin{align}
D^{I}\equiv \sqrt{3}D_L^8+2D_R^3&= |\RED{c_1}|^2 - |\RED{d_1}|^2 + |\RED{e_2}|^2 - |\RED{f_2}|^2 + 2 |\RED{e_3}|^2 - 2 |\RED{f_3}|^2 - |\RED{e_4}|^2 + |\RED{f_4}|^2,\label{equation:D-first}\\
D^{II}\equiv \phantom{\sqrt{3}D_L^8}-2 D_R^3&= |\RED{c_2}|^2 - |\RED{d_2}|^2 + |\RED{e_2}|^2 - |\RED{f_2}|^2 + 2 |\RED{e_1}|^2 - 2 |\RED{f_1}|^2 - |\RED{e_5}|^2 + |\RED{f_5}|^2,\\
D^{III}\equiv \phantom{\sqrt{3}}D_R^6+iD_R^7&=
\phantom{+}\RED{c_1} \RED{c_2}^\ast - \RED{d_1}^\ast \RED{d_2} + \sqrt{2} \RED{e_1}^\ast \RED{e_2} - \sqrt{2} \RED{f_1} \RED{f_2}^\ast\nonumber\\
&\quad+ \sqrt{2} \RED{e_2}^\ast \RED{e_3} - \sqrt{2} \RED{f_2} \RED{f_3}^\ast + \RED{e_4}^\ast \RED{e_5} -
 \RED{f_4} \RED{f_5}^\ast.\label{equation:D-last}
\end{align}
\noindent
The term $D^{III}$ is complex, so it represents $2$ real equations.

\subsection{Symmetries and the general solving strategy}

Before proceeding to solve the equations of motion, it is instructive to note two types of symmetry they possess. These symmetries will have implications on the general strategy, how to solve these equations.

\begin{enumerate}
\item
\textit{Conjugation symmetry}: the Higgs sector contains representations in complex conjugate pairs. This symmetry exchanges between the representation and its conjugate, e.g. \hbox{$27\leftrightarrow\overline{27}$} and \hbox{$351'\leftrightarrow\overline{351'}$.} But since the superpotential contains asymmetric invariants  with respect to this symmetry (the cubic invariants for example), we also have to exchange the parameters in front of the invariants. Explicitly, conjugation symmetry can be written as
\begin{align}
    c_i&\leftrightarrow d_i,\\
    e_i&\leftrightarrow f_i,\\
    \lambda_1&\leftrightarrow \lambda_2,\\
    \lambda_3&\leftrightarrow \lambda_4,\\
    \lambda_5&\leftrightarrow \lambda_6.
\end{align}

Under this symmetry operations, the superpotential $W$ is invariant, so the $F$-terms do not change, while the $D$-terms change according to
$D^{I}\mapsto -D^{I}$, $D^{II}\mapsto -D^{II}$, $D^{III}\mapsto -D^{III\ast}$, so we get an equivalent set of equations of motion.

The exchange of parameters $\lambda$ in front of invariants under conjugation of representations will have major consequences on our strategy of solving the equations of motion. In the absence of this feature, we could start with an ansatz \hbox{$\langle 351'\rangle=\langle\overline{351'}\rangle$} and \hbox{$\langle 27\rangle=\langle \overline{27}\rangle$} (more specifically $c_i=d_i$ and $e_i=f_i$), which automatically solves the $D$-terms, and then only the $F$-terms would remain. But due to the exchange in $\lambda$'s, this ansatz leads to a consistent set of $F$-terms only if the VEVs vanish or we make an exact fine-tuning $\lambda_1=\lambda_2$ and $\lambda_3=\lambda_4$. Since we would like to avoid relations among parameters altogether, at least for symmetry breaking, we abandon this route. The $D$-terms will have to be solved in a non-trivial way, which is not \textit{a priori} obvious, so we find the best strategy to first solve the $F$-terms, and only then proceed to the $D$-terms.


\item
\textit{Alignment symmetry}: by examining the equations of motion, it is also possible to see a symmetry under the exchanges
 \begin{align}
 c_1&\leftrightarrow c_2,&d_1&\leftrightarrow d_2,\label{AS-first}\\
 e_1&\leftrightarrow e_3,&f_1&\leftrightarrow f_3,\\
 e_4&\leftrightarrow e_5,&f_4&\leftrightarrow f_5\label{AS-last}.
 \end{align}
The superpotential $W$ remains unchanged under these exchanges, since every single invariant remains unchanged. Furthermore, the $D$-terms are exchanged according to $D^{I}\leftrightarrow D^{II}$ and $D^{III}\leftrightarrow D^{III\ast}$. All equations of motion thus remain unchanged.

When performing the alignment symmetry operation, we are in fact exchanging the two $\bar{5}$'s of $\SU(5)$ in the representation $27$. By doing this, we are also changing the way $\SO(10)$ and its subgroups (such as Pati-Salam) are embedded in $E_6$, while still containing the same SM group. To elucidate this argument further, consider the $E_6$ subgroup $\SU(2)_R'$ defined by the generators $t_R^6$, $t_R^7$ and $t_R^3-\sqrt{3} t_R^8$ (these are all SM singlets). This $\SU(2)_R'$ is a subgroup of $\SU(3)_R$ in $E_6$, which rotates the second and third component in the $3$ of $\SU(3)_R$. The Standard Model generators, and even $\SU(5)$ generators, commute with $\SU(2)_R'$ rotations, therefore ensuring that the $\SU(2)_R'$ rotations do not change the embedding of either $\SU(5)$ or the SM into $E_6$. But $\SU(2)_R'$ rotations do not commute with the standard $\SU(2)_R$ in $\SU(3)_R$, so the embedding of $\SU(2)_R$ is changed. Since both Pati-Salam and $\SO(10)$ contain the $\SU(2)_R$, the embedding of these two groups also changes.

\end{enumerate}


\subsection{The main branch of solutions}

In accordance with the discussion on conjugation symmetry, we first start by solving the $F$-terms obtained from the superpotential in equation~\eqref{equation-superpotential}. Since the superpotential is renormalizable and its highest order terms are cubic, we get a holomorphic system of $14$ quadratic polynomial equations containing $14$ variables. The general strategy of solving consists of finding equations with a variable only in a linear term, so that we can express this variable from the equation in a unique way.

There are two main branches of solutions, which partly overlap. The first branch conforms to the assumptions $c_1,d_1,e_5,f_5 \neq 0$, while the second branch  assumes $c_2,d_2,e_4,f_4\neq 0$. The two branches are main branches in the sense that the assumptions of nonzero VEVs are general, while zero VEVs would be considered a special kind of ansatz. The assumptions of the two main branches are exchanged under alignment symmetry; this symmetry also brings one main branch into the other, so we will limit our discussion to the first branch.

First, it is possible to express $e_1$, $f_1$, $e_3$, $f_3$ from the terms $F_{f_1}$, $F_{e_1}$, $F_{f_3}$ and $F_{e_3}$, respectively.
To proceed along the first branch, we then express $e_2$ and $f_2$ from $F_{d_2}$ and $F_{c_2}$ respectively, where the assumptions $c_1,d_1\neq 0$ are needed, since $c_1$ and $d_1$ come into the denominators of expressions. Next, we determine $e_4$ and $f_4$ from equations $F_{e_2},F_{f_2}$, respectively, where
the assumption $e_5,f_5\neq 0$ is needed for the same reason. With this procedure, solving the remaining $F$-terms also for $d_1$ and $f_5$, we get the analytic ansatz of the first branch, which solves all the $F$-terms:

\begin{align}
    d_1&=\frac{m_{351'} m_{27}-2 \lambda_3 \lambda_4 \RED{c_2} \RED{d_2}}{2 \lambda_3 \lambda_4 \RED{c_1}},\label{general-first}\\
    e_1&=-\frac{\lambda_3 \RED{c_2}^2+\frac{m_{351'}^2 (m_{351'} m_{27}-2 \lambda_3 \lambda_4 \RED{c_2} \RED{d_2})^2}{108 m_{27}^2 \lambda_1^2 \lambda_2 \RED{e_5}^2}}{m_{351'}},\\
    f_1&=-\frac{\lambda_4 \RED{d_2}^2+3 \lambda_1 \RED{e_5}^2}{m_{351'}},\\
    e_2&=\frac{\lambda_3 \RED{c_1} \left(m_{27} \lambda_4 \RED{d_2} m_{351'}^3-2 \lambda_3 \lambda_4^2 \RED{c_2} \RED{d_2}^2 m_{351'}^2-54 m_{27}^2 \lambda_1^2 \lambda_2 \RED{c_2} \RED{e_5}^2\right)}{27 \sqrt{2} m_{351'} m_{27}^2 \lambda_1^2 \lambda_2 \RED{e_5}^2},\\
    f_2&=\frac{2 \lambda_3 \RED{c_2} \left(\lambda_4 \RED{d_2}^2+3 \lambda_1 \RED{e_5}^2\right)-m_{351'} m_{27} \RED{d_2}}{\sqrt{2} m_{351'} \lambda_3 \RED{c_1}},\\
    e_3&=\frac{\lambda_3 \RED{c_1}^2 \left(-\frac{m_{351'}^2 \lambda_3 \lambda_4^2 \RED{d_2}^2}{m_{27}^2 \lambda_1^2 \lambda_2 \RED{e_5}^2}-27\right)}{27 m_{351'}},\\
    f_3&=-\frac{m_{351'}^2 m_{27}^2-4 m_{351'} \lambda_3 \lambda_4 \RED{c_2} \RED{d_2} m_{27}+4 \lambda_3^2 \lambda_4 \RED{c_2}^2 \left(\lambda_4 \RED{d_2}^2+3 \lambda_1 \RED{e_5}^2\right)}{4 m_{351'} \lambda_3^2 \lambda_4 \RED{c_1}^2},\\
    e_4&=\frac{\RED{c_2} \RED{e_5}}{\RED{c_1}},\\
    f_4&=\frac{m_{351'} \lambda_3 \lambda_4 \RED{c_1} \RED{d_2}}{9 m_{27} \lambda_1 \lambda_2 \RED{e_5}},\\
    f_5&=\frac{m_{351'} (m_{351'} m_{27}-2 \lambda_3 \lambda_4 \RED{c_2} \RED{d_2})}{18 m_{27} \lambda_1 \lambda_2 \RED{e_5}}.\label{general-last}
\end{align}

The VEVs $c_1$, $c_2$, $d_2$ and $e_5$ remain undetermined in the expressions of the ansatz and we use them as variables for the remaining VEVs. They can be determined by solving the $D$-term equations~\eqref{equation:D-first}-\eqref{equation:D-last} with the above ansatz plugged-in. Obtaining all the possible solutions in the branch would involve solving a very complicated system of non-holomorphic polynomials; but a simple solution does exist, if we assume the ansatz $c_2=d_2=0$: equation $D^{III}$ is then solved trivially, while equation $D^{II}$ determines $e_5$. We get a specific solution

\begin{align}
    c_2&=0,&d_2&=0,\label{equation:specific-first}\\
    e_2&=0,&f_2&=0,\\
    e_4&=0,&f_4&=0,\\
    &&d_1&=\frac{m_{351'} m_{27}}{2 \lambda_3 \lambda_4 \RED{c_1}},\\
    e_1&=-\frac{m_{351'}}{6 \lambda_1^{2/3} \lambda_2^{1/3}},&f_1&=-\frac{m_{351'}}{6 \lambda_1^{1/3} \lambda_2^{2/3}},\\
    e_3&=-\lambda_3 \RED{c_1}^2/m_{351'},&f_3&=-\frac{m_{351'} m_{27}^2}{4 \lambda_3^2 \lambda_4 \RED{c_1}^2},\\
    e_5&= \frac{m_{351'}}{3 \sqrt{2} \lambda_1^{2/3} \lambda_2^{1/3}},&f_5&=\frac{m_{351'}}{3 \sqrt{2} \lambda_1^{1/3} \lambda_2^{2/3}}\label{equation:specific-last}.
\end{align}

The term $D^I$ becomes a polynomial condition for $|c_1|^2$:
\begin{align}
    0&= |m_{351'}|^4 |m_{27}|^4 + 2 |m_{351'}|^4 |m_{27}|^2 |\lambda_3|^2 |\RED{c_1}|^2\nonumber\\
    &\qquad -8 |m_{351'}|^2 |\lambda_3|^4 |\lambda_4|^2 |\RED{c_1}|^6 - 16 |\lambda_3|^6 |\lambda_4|^2 |\RED{c_1}|^8.\label{equation:c1-polynomial}
\end{align}
Note that the positive constant and negative coefficient in front of the highest power of $|c_1|$ ensure, that this polynomial always has a positive solution for $|c_1|$. The explicit form of $c_1$ will not be needed.

We get the other main branch of solutions, if we perform the alignment symmetry operation on the ansatz for the first main branch. We also get a specific solution to the $D$-terms from the specific solution of the first branch by alignment symmetry; this is equivalent to a $90^\circ$ real rotation by $\SU(2)_R'$, which brings the second entry of the $3$ of $\SU(3)_R$ to the third entry. There also exists a specific solution, which corresponds to a $45^\circ$ $\SU(2)_R'$ rotation: we get it by the alignment symmetric ansatz $c_1=d_1$, $c_2=d_2$, $e_1=e_3$, $f_1=f_3$, $e_4=e_5$, $f_4=f_5$. This alignment symmetric solution has all VEVs nonzero and is in the overlap of the two main branches.

The solutions in the main branches are in a sense equivalent, since choosing one or the other essentially means choosing, which combination of the $\overline{5}$'s in the $27$ is the Standard Model $\overline{5}$.

To show that the specific solution really breaks to the SM, with no flat directions, we explicitly compute the masses  of the gauge bosons and
of the SM singlets. Everything checks out OK, with further details provided in Appendix~\ref{Appendix:vacuum}.

\subsection{Discussion of alternative solutions\label{section:alternative-solutions}}

Beside the main branches, there are numerous other possible solutions to the equations of motion, which we get by carefully avoiding the assumptions of the two main branches. But it turns out all other solutions of the equations of motion do not break to the SM group, so the above branches are the only solutions for a direct $E_6$ breaking.

In fact, all but one of the alternative solutions leave $\SU(5)$ unbroken. The exception is the solution with the ansazt $\langle 27\rangle=\langle \overline{27}\rangle=0$: this case corresponds to the model with breaking sector $351'+\overline{351'}$ in section~\ref{section:351'-model}. Due to the presence of the cubic invariants, this is again an example of a model where the ansatz $\langle 351'\rangle=\langle\overline{351'}\rangle$ is not valid and the $D$-terms need to be solved nontrivially.

Solving the $F$-terms in the case $\langle 27\rangle=\langle \overline{27}\rangle=0$ is a simple matter:
 \begin{align}
    c_1&=0,&\label{PS-ansatz-begin}
    d_1&=0,\\
    c_2&=0,&
    d_2&=0,\\
    e_1&=-\frac{3 \lambda_2 \RED{f_5}^2}{m_{351'}},&
    f_1&=-\frac{3 \lambda_1 \RED{e_5}^2}{m_{351'}},\\
    e_2&=\frac{3 \sqrt{2} \lambda_2 \RED{f_4} \RED{f_5}}{m_{351'}},&
    f_2&=\frac{3 \sqrt{2} \lambda_1 \RED{e_4} \RED{e_5}}{m_{351'}},\\
    e_3&=-\frac{3 \lambda_2 \RED{f_4}^2}{m_{351'}},&
    f_3&=-\frac{3 \lambda_1 \RED{e_4}^2}{m_{351'}},\\
   m_{351'}^2&=18 \lambda_1 \lambda_2( \RED{e_4} \RED{f_4} + \RED{e_5} \RED{f_5}).\label{PS-ansatz-end}
 \end{align}

By explicit computation we discover 21 massless gauge bosons before even solving for the $D$-terms. This scenario thus breaks to a Pati-Salam group, which is in general embedded into $E_6$ in a (possibly) non-canonical way. The canonical embedding is recovered by
the ansatz $e_5=f_5=0$, which properly aligns the invariant Pati-Salam. In fact the only nonzero VEVs are then $e_3$, $f_3$, $e_4$ and $f_4$, exactly the ones which are singlets under the canonical Pati-Salam (see Table~\ref{table-singlets}).

It was therefore the addition of the VEVs from $27+\overline{27}$, which enabled $E_6$  to be broken all the way to the SM.

\section{Where are the MSSM Higgses?\label{section:MSSM-Higgses}}

The MSSM Higgses would be expected to reside in the breaking sector \hbox{$27+\overline{27}+351'+\overline{351'}$.} More specifically, they would need to reside at least partly in representations, which couple to the fermionic pair $27_F^i\, 27_F^j$, so they should be present in $27$ and $\overline{351'}$. They cannot reside in $27_F$, since there is no cubic term $27_F^3$ due to matter parity.

The usual procedure would be to compute the mass matrices of the doublets $(1,2,+\tfrac{1}{2})$ and antidoublets $(1,2,-\tfrac{1}{2})$ and perform a fine-tuning of Lagrangian parameters to get one doublet mode massless; the soft supersymmetry-breaking terms would then enable this mode to get an electroweak (EW) scale VEV. At the same time, we need to make sure the fine-tuning still keeps the triplets $(3,1,-\tfrac{1}{3})$ and antitriplets $(\overline{3},1,+\tfrac{1}{3})$ heavy (of the order of the GUT scale), since they mediate proton decay. This separation of scales is called the doublet-triplet (DT) splitting problem. Although fine-tuning is not considered to be aesthetically pleasing, it usually does the job.

But curiously, in our case, DT splitting cannot be performed: all vacua, which break to the SM, have this inability, since making the doublet massless automatically does the same to the triplet. Further computational details on the this DT splitting attempt in the breaking sector are provided in Appendix~\ref{Original-DT-splitting}, as well as a list of possible reasons for failure. The inability to perform DT splitting is disappointing, since it is only this usually trivial hurdle which prevents the $27+\overline{27}+351'+\overline{351'}$ model to be realistic.

The above DT problem can be cured by a reasonably simple action: we add an extra pair $\widetilde{27}+\overline{\widetilde{27}}$ to the model.
To simplify the analysis we will assume that these extra  $\widetilde{27}+\overline{\widetilde{27}}$
couple only quadratically with the Higgs fields with large VEVs. In this way we are making
automatic the solution of the old equations of motion for vanishing VEVs of these new tilde fields. In addition, the mass matrices of the doublets and triplets in the tilde fields decouple from the DT matrices of the breaking sector.

We thus get the following tilde-field superpotential needed for the doublet-triplet splitting:

\begin{align}
W_{DT}&=m_{\widetilde{27}}\;\widetilde{27}\;\overline{\widetilde{27}}+\kappa_1\;\widetilde{27}\;\widetilde{27}\;\overline{351'}+\kappa_2\;\overline{\widetilde{27}}\;\overline{\widetilde{27}}\;351'
+\kappa_3 \; \widetilde{27}\;\widetilde{27}\;27+\kappa_4 \; \overline{\widetilde{27}}\;\overline{\widetilde{27}}\;\overline{{27}}.
\end{align}

The tilde sector contains $3$ doublet/antidoublet and triplet/antitriplet pairs, so the mass matrices will be $3\times 3$. For our labeling convention and list of the doublets and triplets in the tilde representations, see Table~\ref{table:doublets}.  The mass matrices can be written as

\begin{align}
\widetilde{\mathcal{M}}_{\textrm{doublets}}&=
\begin{pmatrix}
 m_{\widetilde{27}} & -2 \kappa_3 \RED{c_1}-3\kappa_1 \frac{\RED{f_4}}{\sqrt{15}} & 2 \kappa_3\RED{c_2}-3 \kappa_1 \frac{\RED{f_5}}{\sqrt{15}} \\
 -2 \kappa_4 \RED{d_1}-3 \kappa_2 \frac{\RED{e_4}}{\sqrt{15}} & m_{\widetilde{27}} & 0 \\
 \phantom{-}2 \kappa_4 \RED{d_2}-3  \kappa_2 \frac{\RED{e_5}}{\sqrt{15}} & 0 & m_{\widetilde{27}}\\
\end{pmatrix},\\
\widetilde{\mathcal{M}}_{\textrm{triplets}}&=
\begin{pmatrix}
m_{\widetilde{27}} & -2 \kappa_3\RED{c_1}+2\kappa_1 \frac{\RED{f_4}}{\sqrt{15}} & 2 \kappa_3 \RED{c_2}+2 \kappa_1 \frac{\RED{f_5}}{\sqrt{15}} \\
 -2 \kappa_4 \RED{d_1}+2 \kappa_2 \frac{\RED{e_4}}{\sqrt{15}} & m_{\widetilde{27}} & 0 \\
 \phantom{-}2 \kappa_4 \RED{d_2}+2 \kappa_2 \frac{\RED{e_5}}{\sqrt{15}} & 0 & m_{\widetilde{27}}\\
\end{pmatrix},\label{equation:triplet-mass-tilde}
\end{align}

\noindent
where the mass terms are written as

\begin{align}
\begin{pmatrix}
\widetilde{D}_1&\widetilde{D}_2&\widetilde{D}_3\\
\end{pmatrix}
\;\widetilde{\mathcal{M}}_{{\rm doublets}}\;
\begin{pmatrix}
\overline{\widetilde{D}}_1\\
\overline{\widetilde{D}}_2\\
\overline{\widetilde{D}}_3\\
\end{pmatrix}
+\begin{pmatrix}
\widetilde{T}_1&\widetilde{T}_2&\widetilde{T}_3 \\
\end{pmatrix}
\;\widetilde{\mathcal{M}}_{{\rm triplets}}\;
\begin{pmatrix}
\overline{\widetilde{T}}_1\\
\overline{\widetilde{T}}_2\\
\overline{\widetilde{T}}_3\\
\end{pmatrix}.
\end{align}

\begin{table}[h!]
\caption{Labels of the doublets and triplets along with their locations in $\widetilde{27}$ and $\overline{\widetilde{27}}$. The corresponding EW-VEVs are also labeled.\label{table:doublets}}
\vskip 0.3cm
\centering
\begin{tabular}{lllll}
\toprule
doublet,triplet&$\subset\SU(5)$&$\subset\SO(10)$&$\subset E_6$&doublet VEV\\\midrule
$\widetilde{D}_1,\widetilde{T}_1$&$5$&$10$&$\widetilde{27}$&$v_1$\\
$\widetilde{D}_2,\widetilde{T}_2$&$5$&$10$&$\overline{\widetilde{27}}$&$v_2$\\
$\widetilde{D}_3,\widetilde{T}_3$&$5$&$\overline{16}$&$\overline{\widetilde{27}}$&$v_3$\\\addlinespace
$\overline{\widetilde{D}}_1,\overline{\widetilde{T}}_1$&$\overline{5}$&$10$&$\overline{\widetilde{27}}$&$\bar v_1$\\
$\overline{\widetilde{D}}_2,\overline{\widetilde{T}}_2$&$\overline{5}$&$10$&$\widetilde{27}$&$\bar v_2$\\
$\overline{\widetilde{D}}_3,\overline{\widetilde{T}}_3$&$\overline{5}$&$16$&$\widetilde{27}$&$\bar v_3$\\\bottomrule
\end{tabular}
\end{table}

A fine tuning among the new $\kappa$ parameters will ensure DT splitting. If we plug the vacuum solution into the mass matrices $\widetilde{\mathcal{M}}_{\textrm{doublets}}$ and $\widetilde{\mathcal{M}}_{\textrm{triplets}}$, we get the following DT splitting conditions:
\begin{align}
0&= m_{\widetilde{27}}^3 - \frac{1}{30}\;\;m_{\widetilde{27}} m_{351'}^2 \frac{\kappa_1 \kappa_2}{\lambda_1 \lambda_2} - 2 m_{\widetilde{27}} m_{351'} m_{27} \frac{\kappa_3 \kappa_4}{\lambda_3 \lambda_4},\\
0&\neq m_{\widetilde{27}}^3 - \frac{2}{135}\; m_{\widetilde{27}} m_{351'}^2 \frac{\kappa_1 \kappa_2}{\lambda_1 \lambda_2} - 2 m_{\widetilde{27}} m_{351'} m_{27} \frac{\kappa_3 \kappa_4}{\lambda_3 \lambda_4}.
\end{align}

\noindent
These two conditions insure a massless doublet mode, but keep all triplet modes heavy. Both can be simultaneously satisfied by a fine-tuning
\begin{align}
\kappa_1 &\approx 30\,(m_{\widetilde{27}}^2 \lambda_3 \lambda_4 - 2 m_{351'} m_{27} \kappa_3 \kappa_4)\frac{\lambda_1 \lambda_2 }{m_{351'}^2 \lambda_3 \lambda_4 \kappa_2}.\label{equation:kappa1-fine-tuning}
\end{align}

The above fine-tuning of $\kappa_1$ gives the following modes of doublets and antidoublets to be massless:
\begin{align}
\widetilde{D}_{m=0}&\propto\phantom{+}\frac{\sqrt{1/30}\;m_{\widetilde{27}} m_{351'} \lambda_1^{-2/3} \lambda_2^{-1/3} \lambda_3 \lambda_4 \kappa_2}{
 m_{\widetilde{27}}^2 \lambda_3 \lambda_4 -
    2 m_{351'} m_{27} \kappa_3 \kappa_4}\;\widetilde{D}_1 \nonumber\\
 &\quad+ \frac{\sqrt{2/15}\;m_{351'}\RED{c_1} \lambda_1^{-2/3} \lambda_2^{-1/3}\lambda_3 \lambda_4 \kappa_2 \kappa_3}{m_{\widetilde{27}}^2 \lambda_3 \lambda_4 - 2 m_{351'} m_{27} \kappa_3 \kappa_4} \;\widetilde{D}_2 + \widetilde{D}_3,\\
\overline{\widetilde{D}}_{m=0}&\propto\frac{\sqrt{30}\;m_{\widetilde{27}} \lambda_1^{2/3} \lambda_2^{1/3}}{m_{351'} \kappa_2}\;\overline{\widetilde{D}}_1+ \frac{\sqrt{30}\; m_{27} \lambda_1^{2/3} \lambda_2^{1/3} \kappa_4}{\RED{c_1} \lambda_3 \lambda_4 \kappa_2}\;\overline{\widetilde{D}}_2+\overline{\widetilde{D}}_3.
\end{align}

Notice that the massless modes have components of all doublets and antidoublets present: in particular, the Higgs is present in the components $\widetilde{D}_1$, $\overline{\widetilde{D}}_2$ and $\overline{\widetilde{D}}_3$  of $\widetilde{27}$, so the corresponding EW VEVs $v_1$, $\bar{v}_2$ and $\bar{v}_3$ all become nonzero. The presence of these VEVs will be important in the analysis of the Yukawa sector.

\section{The Yukawa sector\label{section:Yukawa-sector}}

Assuming a $\mathbb{Z}_2$ ``matter parity'', which avoids potentially dangerous R-parity violating terms, with
\begin{align}
27_F&\to-27_F
\end{align}
\noi
and with all other fields even, we can write down the most general Yukawa sector:
\begin{align}
\mathcal{L}_{\textrm{Yukawa --}E_6}= \tfrac{1}{2}\;27_F^i\;27^j_F\;\left(\yd^{ij}\;27+\yt^{ij}\;\overline{351'}+\ydt^{ij}\;\widetilde{27}\right).
\end{align}
Excluding the tilde part, our $E_6$ Yukawa sector is completely analogous to the Yukawa sector in the minimal renormalizable $\SO(10)$ model \cite{Babu:1992ia,Bajc:2002iw}:

\begin{align}
\mathcal{L}_{\textrm{Yukawa --}\SO(10)}= \tfrac{1}{2}\;16_F^i\;16^j_F\;\left(Y_{10}^{ij}\;10+Y^{ij}_{\overline{126}}\;\overline{126}\right).\phantom{+Y_{27}\;27'}
\end{align}

Our $27$ plays the role of the $16$, and our $\overline{351'}$ plays the role of $\overline{126}$. Since $10\subset 27$ and $\overline{126}\subset\overline{351'}$, our terms include the terms from the renormalizable $\SO(10)$ case, as well as some additional terms like
$16^F_i 10^F_i (Y^{27}\; 16+Y^{\overline{351'}}\; 144)$.

Notice, however, that the flavor-mixing mechanism in the two cases is completely different. In the renormalizable $\SO(10)$ we have the usual case of GUTs where the EW Higgs is present in two representations: $10$ and $\overline{126}$. Since the two generic matrices $Y_{10}$ and $Y_{\overline{126}}$ cannot be diagonalized simultaneously, we get flavor mixing.

Flavor mixing in our model is more subtle. The MSSM Higgs doublets are present only in  $\widetilde{27}$ (and $\overline{\widetilde{27}}$, which is not present in the Yukawa sector), but the representations
$27$ and $\overline{351'}$ of the breaking sector acquire GUT scale VEVs. Through these large $\SU(5)$ breaking VEVs, they contribute to mix the $\bar 5$ in $16$ with the $\bar 5$ in $10$ of $27$. Flavor mixing is therefore not due to the presence of Higgs in two different representations at the EW scale, but due to the different mixing of vector-like heavy pairs at the GUT scale. This situation is analogous to \cite{Barbieri:1980vc}, which we further elaborate on below.

The mass matrices are explicitly computed to be (we skip their hermitian conjugate part)

\begin{align}
u^{T}(-\BLUE{v_1})\ydt u^c +
\bem
d^{cT} & d'^{cT}\\
\eem
\bem
\phantom{-}\BLUE{\bar{v}_2}\ydt & \phantom{-}\RED{c_2}\yd+\frac{\RED{f_5}}{\sqrt{15}}\yt \\
-\BLUE{\bar{v}_3}\ydt & -\RED{c_1}\yd+\frac{\RED{f_4}}{\sqrt{15}}\yt \\
\eem\!\!
\bem
d \\
d'\\
\eem \nonumber\\
+
\bem
e^{T} & e'^{T}
\eem
\bem
-\BLUE{\bar{v}_2}\ydt & \phantom{-}\RED{c_2}\yd-\frac{3}{2}\frac{\RED{f_5}}{\sqrt{15}}\yt \\
\phantom{-}\BLUE{\bar{v}_3}\ydt & -\RED{c_1}\yd-\frac{3}{2}\frac{\RED{f_4}}{\sqrt{15}}\yt\\
\eem\!\!
\bem
e^c \\
e'^c\\
\eem &&\nonumber\\
+
\bem
\nu^{T} & \nu'^{T}\\
\eem \!
\bem
\BLUE{v_1}\ydt & 0 & \phantom{-}\RED{c_2}\yd-\frac{3}{2}\frac{\RED{f_5}}{\sqrt{15}}\yt \\
0 & -\BLUE{v_1}\ydt & -\RED{c_1}\yd-\frac{3}{2}\frac{\RED{f_4}}{\sqrt{15}}\yt\\
\eem \!\!
\bem
\nu^c \\
s \\
\nu'^c
\eem \nonumber\\
+\frac{1}{2}
\bem
\nu^{cT} & s^T & \nu'^{cT}\\
\eem\!
\bem
\RED{f_1}\yt & \frac{\RED{f_2}}{\sqrt{2}}\yt & -\BLUE{\bar{v}_3}\ydt \\
\frac{\RED{f_2}}{\sqrt{2}}\yt & \RED{f_3}\yt & \phantom{-}\BLUE{\bar{v}_2}\ydt \\
-\BLUE{\bar{v}_3}\ydt & \BLUE{\bar{v}_2}\ydt & 0\\
\eem\!\!
\bem
\nu^c \\
s \\
\nu'^c\\
\eem\nonumber\\
+\frac{1}{2}
\begin{pmatrix}
\nu^T & \nu'^T
\end{pmatrix}
\begin{pmatrix}
\Delta_1 Y_{\overline{351}'} & \tfrac{1}{\sqrt{2}}\Delta_2 Y_{\overline{351}'}\\
\tfrac{1}{\sqrt{2}}\Delta_2 Y_{\overline{351}'}& \Delta_3 Y_{\overline{351}'}\\
\end{pmatrix}
\begin{pmatrix}
\nu \\
\nu' \\
\end{pmatrix}\!.
\end{align}

\noi
where the barred $\bar\Delta\sim (1,3,+1)$ and unbarred $\Delta\sim(1,3,-1)$ weak triplets
shown in Table~\ref{table:pn-weak-triplets} contribute to the type II seesaw.

\begin{table}[h!]
\caption{Labels for weak triplets $(1,3,\pm 1)$ relevant for seesaw type II.\label{table:pn-weak-triplets}}
\vskip 0.2cm
\centering
\begin{tabular}{llp{1cm}@{\hspace{2cm}}llp{1cm}}
\toprule
label&{\footnotesize $E_6\supseteq\SO(10)\supseteq\SU(5)$}&p.n.&label&{\footnotesize $E_6\supseteq\SO(10)\supseteq\SU(5)$}&p.n.\\\midrule
$\overline{\Delta}_1$&$351'\supseteq 126\supseteq\phantom{0}\overline{15}$&$L\phantom{'}L$&$\Delta_1$&$\overline{351'}\supseteq\overline{126}\supseteq 15$&$\bar{L}\phantom{'}\bar{L}$\\
$\overline{\Delta}_2$&$351'\supseteq\overline{144}\supseteq\phantom{0}\overline{15}$&$L\phantom{'}L'$&$\Delta_2$&$\overline{351'}\supseteq 144\supseteq 15$&$\bar{L}\phantom{'}\bar{L}'$\\
$\overline{\Delta}_3$&$351'\supseteq\phantom{0}54\supseteq\phantom{0}\overline{15}$&$L'L'$&$\Delta_3$&$\overline{351'}\supseteq\phantom{0}54\supseteq 15$&$\bar{L}'\bar{L}'$\\
$\Delta_4$&$351'\supseteq\phantom{0}54\supseteq\phantom{0}15$&$L'^c L'^c$&$\overline{\Delta}_4$&$\overline{351'}\supseteq\phantom{0}54\supseteq\overline{15}$&$\bar{L}'^c \bar{L}'^c$\\\bottomrule
\end{tabular}
\end{table}

The triplets get nonzero VEVs as usual: the superpotential terms are

\begin{align}
W\big|_{\textrm{triplets}}&=
\begin{pmatrix}
\overline{\Delta}_1&\overline{\Delta}_2&\overline{\Delta}_3&\overline{\Delta}_4\\
\end{pmatrix}
\begin{pmatrix}
 m_{351'} & 0 & 0 & \phantom{-}6 \lambda_1\RED{e_1} \\
 0 & m_{351'} & 0 & -6 \lambda_1\RED{e_2} \\
 0 & 0 & m_{351'} & \phantom{-}6\lambda_1\RED{e_3} \\
 6 \lambda_2\RED{f_1} & -6\lambda_2\RED{f_2} & 6\lambda_2\RED{f_3} & \phantom{-}m_{351'}\\
\end{pmatrix}
\begin{pmatrix}
\Delta_1\\ \Delta_2\\\Delta_3\\\Delta_4\\
\end{pmatrix}\nonumber\\
&\qquad +
\begin{pmatrix}
\overline{\Delta}_1&\overline{\Delta}_2&\overline{\Delta}_3&\overline{\Delta}_4\\
\end{pmatrix}
\begin{pmatrix}
\kappa_2 \BLUE{v_3}^2\\ \kappa_2 \sqrt{2}\BLUE{v_2} \BLUE{v_3}\\ \kappa_2 \BLUE{v_2}^2\\ \kappa_1 \BLUE{v_1}^2\\
\end{pmatrix}\nonumber\\
&\qquad +
\begin{pmatrix}
\kappa_1\BLUE{\overline{v}_3}^2&\kappa_1\sqrt{2}\BLUE{\overline{v}_3}\BLUE{\overline{v}_2}&\kappa_1\BLUE{\overline{v}_2}^2&\kappa_2\BLUE{\overline{v}_1}^2
\end{pmatrix}
\begin{pmatrix}
\Delta_1\\ \Delta_2\\\Delta_3\\\Delta_4\\
\end{pmatrix}.
\end{align}
Integrating out the heavy triplets yields
\begin{align}
\begin{pmatrix}
\Delta_1\\ \Delta_2\\\Delta_3\\\Delta_4\\
\end{pmatrix}&=
\begin{pmatrix}
 m_{351'} & 0 & 0 & \phantom{-}6 \lambda_1\RED{e_1} \\
 0 & m_{351'} & 0 & -6 \lambda_1\RED{e_2} \\
 0 & 0 & m_{351'} & \phantom{-}6\lambda_1\RED{e_3} \\
 6 \lambda_2\RED{f_1} & -6\lambda_2\RED{f_2} & 6\lambda_2\RED{f_3} & \phantom{-}m_{351'}\\
\end{pmatrix}^{-1}
\begin{pmatrix}
\kappa_2 \BLUE{v_3}^2\\ \kappa_2 \sqrt{2}\BLUE{v_2} \BLUE{v_3}\\ \kappa_2 \BLUE{v_2}^2\\ \kappa_1 \BLUE{v_1}^2\\
\end{pmatrix}.\label{delta-triplets}
\end{align}

To get the light fermion mass matrices explicitly, we will integrate out the
heavy vector-like pairs following for example \cite{Barr:2003zx,Babu:2012pb}.
Our matrices are of block form

\begin{align}
{\cal M}=
\begin{pmatrix}
M_1 & A \\
M_2 & B \\
\end{pmatrix},
\end{align}

\noi
where $M_{1,2}$ are $3\times 3$ matrices of order ${\cal O}(m_{W})$
while $A,B$ are $3\times 3$ matrices of order ${\cal O}(M_{GUT})$. If we multiply
from the left with

\begin{align}
{\cal U}&=
\begin{pmatrix}
\Lambda & -\Lambda X \cr
X^\dagger \Lambda & \bar\Lambda
\end{pmatrix},\label{equation:U-matrix}
\end{align}

\noi
where

\begin{align}
X&=A\,B^{-1},\\
\Lambda&=\left(1+XX^\dagger\right)^{-1/2},\\
\bar\Lambda&=\left(1+X^\dagger X\right)^{-1/2},
\end{align}

\noi
with relations

\bea
X^\dagger \Lambda&=&\bar\Lambda X^\dagger,\\
X\bar\Lambda&=&\Lambda X,
\eea

\noi
we get

\begin{align}
{\cal U\,M}&=
\begin{pmatrix}
\Lambda (M_1-X\,M_2) & 0 \cr
X^\dagger\,\Lambda\,M_1+\bar\Lambda\,M_2 & X^\dagger\,\Lambda\,A+\bar\Lambda\,B
\end{pmatrix}=
\begin{pmatrix}
{\cal O}(m_W) & 0 \\
{\cal O}(m_W) & {\cal O}(M_{GUT})\\
\end{pmatrix}.
\end{align}

Integrating out the heavy states in the lower right part, we are left in leading order of $m_W/M_{GUT}$ with the matrix
for light states
\begin{align}
M&=\Lambda\,(M_1-X\,M_2).
\end{align}
For our solution with the ansatz
\begin{align}
c_2=f_2=f_4&=0,
\end{align}
\noi
and defining
\begin{align}
X_0&\equiv \sqrt{\frac{3}{20}}\,\frac{\RED{f_5}}{\RED{c_1}}\,\yt\yd^{-1},\label{equation:X0-definition}
\end{align}
\noi
the light fermion masses become
\begin{align}
M_D^T&=\left(1+(4/9)X_0X_0^\dagger\right)^{-1/2}\left(\BLUE{\bar{v}_2}-(2/3)\BLUE{\bar{v}_3}X_0\right)\ydt,\\
M_E&=-\left(1+X_0X_0^\dagger\right)^{-1/2}\left(\BLUE{\bar{v}_2}+\BLUE{\bar{v}_3}X_0\right)\ydt,\\
M_U&=-\BLUE{v_1}\ydt,\\
\label{MN}
M_N&=\frac{1}{2}
\left(1+X_0X_0^\dagger\right)^{-1/2}\times\left(\Delta_1Y_{\overline{351}'}-
\frac{\Delta_2}{\sqrt{2}}\left(X_0 Y_{\overline{351}'}+ Y_{\overline{351}'}X_0^T\right)+
\Delta_3 X_0 Y_{\overline{351}'}X_0^T\right.\nonumber\\
&\qquad\left.-\frac{\BLUE{v_1}^2}{\RED{f_1}}\ydt\yt^{-1}\ydt-
\frac{\BLUE{v_1}^2}{\RED{f_3}}X_0\ydt\yt^{-1}\ydt X_0^T\right) \times\left(1+X_0^\ast X_0^T\right)^{-1/2}.
\end{align}

Notice the combined contributions of both type I
\cite{Minkowski:1977sc,Yanagida,gellmannramondslansky,Glashow,Mohapatra:1979ia}
(proportional to $v_1^2$) and type II \cite{Magg:1980ut,Schechter:1980gr,Lazarides:1980nt,Mohapatra:1980yp}
(proportional to $\Delta_{1,2,3}$) seesaw in (\ref{MN}).
As always, the seesaw mechanism gives neutrino masses of the scale $\mathcal{O}(m_{W}^2/M_{GUT})$, which can be seen from the factors $\tfrac{v_1^2}{f_1}$ and $\tfrac{v_1^2}{f_3}$ for type I contributions, while for type II contributions we have $\Delta_i\sim\mathcal{O}(m_{W}^2/M_{GUT})$ from equation~\eqref{delta-triplets}.

There is no type III \cite{Foot:1988aq} seesaw contribution: although there are (fermionic) weak triplets $(1,3,0)$ in the $\overline{351'}$, there is no $27_F\,\widetilde{27}\,\overline{351'}$ term in the superpotential due to the imposed R-parity; in the presence of $R$-parity for $27_F$, a type III seesaw contribution is never possible, since there are no triplets $(1,3,0)$ in the $27_F$.

The general observation on the fermion masses is the following: although the $27_F$ introduces an extra SM singlet (which is another right handed neutrino) and vector-like pairs of quarks and leptons, the extra degrees of freedom are all massive (of the order of $M_{GUT}$). This means we recover the usual low-energy degrees of freedom from the MSSM. Also, the $16$ and $10$ of $\SO(10)$ in the $27_F$'s mix, i.e.~the light states do not live just in the $16$.

The explicit fitting of these mass matrices to the experimental values of the masses and
mixings is complicated by the nonlinear way the various matrices enter into the equations.
This is typical for contributions from vector-like families. Although the full analysis is
beyond the scope of this paper, we note here that there are 3 Yukawa matrices involved.
The number of free parameters seems more than likely large enough to allow a successful fit.
We leave the full analysis for a future publication.

\section{Proton decay\label{section:proton-decay}}

The $D=5$ proton decay \cite{Sakai:1981pk,Weinberg:1981wj,Lucas:1996bc,Goto:1998qg} is mediated
here\footnote{Although $\overline{351'}$ contains
also triplets $(3,1,-\tfrac{4}{3})$, it (as well as the $27$'s) does not contain the antitriplets $(\bar{3},1,\tfrac{4}{3})$. These
are part of $351'$, which however does not couple to the MSSM matter supermultiplets.} by color triplets
of the type $T\sim (3,1,-\tfrac{1}{3})$ and $\overline{T}\sim(\bar{3},1,\tfrac{1}{3})$. All such triplets in our model have been identified in Tables~\ref{table:doublets} and \ref{table:pn-doublets-triplets}: there are $15$ triplet/antitriplet pairs altogether, with $12$ pairs coming from the non-tilde fields $27$, $\overline{27}$, $351'$ and $\overline{351'}$, while $3$ pairs are in the tilde fields $\widetilde{27}$ and $\overline{\widetilde{27}}$. Note that the triplets and antitriplets in $27_F^i$ do not mediate proton decay, since the $\mathbb{Z}_2$ matter parity forbids cubic vertices $27_F^3$.

The full superpotential of our model is
\begin{align}
W_{\textrm{full}}&=m_{27}\; 27\;\overline{27}+m_{351'}\;351'\;\overline{351'}\nonumber+m_{\widetilde{27}}\;\widetilde{27}\;\overline{\widetilde{27}}\\
&\quad+\lambda_1\; 351'^3 + \lambda_2\;\overline{351'}^3+\lambda_3\;27^2\;\overline{351'}+\lambda_4\; \overline{27}^2\;351'+\lambda_5\;27^3+\lambda_6\;\overline{27}^3\nonumber\\
&\quad+\kappa_1\;\widetilde{27}{}^2\;\overline{351'}+\kappa_2\;\overline{\widetilde{27}}{}^2\;351'
+\kappa_3 \; \widetilde{27}{}^2\;27+\kappa_4 \; \overline{\widetilde{27}}{}^2\;\overline{{27}}\nonumber \\
&\quad+\tfrac{1}{2}\yd^{ij}\;27_F^i\;27^j_F\;27+\tfrac{1}{2}\yt^{ij}\;27_F^i\;27^j_F\;\overline{351'}+\tfrac{1}{2}\ydt^{ij}\;27_F^i\;27^j_F\;\widetilde{27}.\label{equation:W-full}
\end{align}

The relevant couplings for proton decay can be written in terms of SM irreducible representations generically as
\begin{align}
\label{Wproton}
W\big|_{\textrm{proton}}&=T_A\,(\mathcal{M}_T)^{AB}\,\overline{T}_B + C_1^{ijA}\; Q_i\,Q_j\,T_A + C_2^{ijA}\; u^c_i\,e^c_j\,T_A \nonumber\\
&\quad+ \overline{C}_1^{ijA}\;Q_i\,L_j\,\overline{T}_A+ \overline{C}_1'^{ijA} Q_i\,L'_j\,\overline{T}_A+\overline{C}_2^{ijA}\;d^c_i\,u^c_j\,\overline{T}_A+\overline{C}_2'^{ijA}\;d'^c_i\,u^c_j\,\overline{T}_A,
\end{align}
where $i,j$ are generation indices and $A,B=1,\ldots,15$ are indices over all the color triplets/antitriplets, with sums over repeated indices; we define $T_{12+A}:=\widetilde{T}_A$ and $\overline{T}_{12+A}:=\overline{\widetilde{T}}_A$ (with $A=1,2,3$). We suppress the $\SU(3)_C$ and $\SU(2)_L$ indices in our notation; the indices in the fields are contracted with the epsilon tensors in the order the fields are written, with $\varepsilon_{123}=\varepsilon_{12}=1$.

The triplet mass matrix $\mathcal{M}_T$ has contributions from the following terms in equation~\eqref{equation:W-full}: the mass terms $m_{27}$, $m_{351'}$ and $m_{\widetilde{27}}$, the $\lambda$-terms and the $\kappa$-terms. The tilde and non-tilde fields do not mix in the mass terms because the tilde fields have vanishing VEVs, so $\mathcal{M}_T$ has the block form
\begin{align}
\mathcal{M}_T
&=
\begin{pmatrix}
\big(\mathcal{M}_{\textrm{triplets}}\big)_{12\times 12}&0\\
0&\big(\widetilde{\mathcal{M}}_{\textrm{triplets}}\big)_{3\times 3}\\
\end{pmatrix}.
\end{align}

The two matrices are
\begin{align}
&\mathcal{M}_{\textrm{triplets}}=\nonumber\\
&\scalebox{0.75}{
$
\hspace{-0.5cm}
\left(
\begin{smallmatrix}
m_{27} & -6 \RED{c_1} \lambda _5 & \frac{\sqrt{2} m_{351'} \lambda _3}{3\sqrt{15} \lambda _1^{1/3} \lambda _2^{2/3}} & -\sqrt{\frac{8}{5}} \RED{c_1} \lambda _3 & 0 & 0 & 0 & 0 & 0 & 0
   & 0 & 0 \\
 -\frac{3 m_{27} m_{351'} \lambda _6}{\RED{c_1} \lambda _3 \lambda _4} & m_{27} & 0 & 0 & -\frac{\sqrt{2} m_{27} m_{351'}}{\sqrt{5}\RED{c_1} \lambda _3} & 0 & 0 & 0 & 0 & 0 & 0 & 0 \\
\frac{\sqrt{2}m_{351'} \lambda _4}{3\sqrt{15}\lambda _1^{2/3} \lambda _2^{1/3}} & 0 & m_{27} & 0 & 0 & -\frac{m_{27} m_{351'}}{\sqrt{2} \RED{c_1} \lambda _3} & 0 & 0 & 0 & 0 & 0 &
   0 \\
 -\frac{\sqrt{2} m_{27} m_{351'}}{\sqrt{5}\RED{c_1} \lambda _3} & 0 & 0 & m_{351'} & 0 & 0 & 0 & 0 & 0 & -\frac{m_{351'} \lambda_1^{1/3}}{\sqrt{6} \lambda _2^{1/3}} & 0 & 0 \\
 0 & -\sqrt{\frac{8}{5}} \RED{c_1} \lambda _3 & 0 & 0 & m_{351'} & 0 & 0 & 0 & -\frac{m_{351'} \lambda_2^{1/3}}{2 \sqrt{30} \lambda_1^{1/3}} & 0 & -\frac{5 m_{351'}
   \lambda_2^{1/3}}{2 \sqrt{6} \lambda_1^{1/3}} & 0 \\
 0 & 0 & -\sqrt{2} \RED{c_1} \lambda _3 & 0 & 0 & m_{351'} & 0 & 0 & 0 & 0 & 0 & 0 \\
 0 & 0 & 0 & 0 & 0 & 0 & m_{351'} & 0 & -\frac{m_{351'} \lambda_2^{1/3}}{2 \sqrt{2} \lambda_1^{1/3}} & 0 & \frac{\sqrt{5} m_{351'} \lambda_2^{1/3}}{6\sqrt{2}
   \lambda_1^{1/3}} & 0 \\
 0 & 0 & 0 & 0 & 0 & 0 & 0 & m_{351'} & 0 & -\frac{\sqrt{5} m_{351'} \lambda_1^{1/3}}{\sqrt{6}\lambda_2^{1/3}} & 0 & 0 \\
 0 & 0 & 0 & 0 & -\frac{m_{351'} \lambda_1^{1/3}}{2 \sqrt{30} \lambda_2^{1/3}} & 0 & -\frac{m_{351'} \lambda_1^{1/3}}{2 \sqrt{2} \lambda_2^{1/3}} & 0 & m_{351'} &
   0 & 0 & 0 \\
 0 & 0 & 0 & -\frac{m_{351'} \lambda_2^{1/3}}{\sqrt{6} \lambda_1^{1/3}} & 0 & 0 & 0 & -\frac{\sqrt{5} m_{351'} \lambda_2^{1/3}}{\sqrt{6}\lambda_1^{1/3}} & 0 &
   m_{351'} & 0 & 0 \\
 0 & 0 & 0 & 0 & -\frac{5 m_{351'} \lambda_1^{1/3}}{2 \sqrt{6} \lambda_2^{1/3}} & 0 & \frac{\sqrt{5} m_{351'} \lambda_1^{1/3}}{6\sqrt{2} \lambda_2^{1/3}} & 0 &
   0 & 0 & m_{351'} & \frac{2 \sqrt{5} m_{351'} \lambda_1^{1/3}}{3 \lambda_2^{1/3}} \\
 0 & 0 & 0 & 0 & 0 & 0 & 0 & 0 & 0 & 0 & \frac{2 \sqrt{5} m_{351'} \lambda_2^{1/3}}{3 \lambda_1^{1/3}} & m_{351'}
\end{smallmatrix}
\right)
$
},
\end{align}

\begin{align}
\widetilde{\mathcal{M}}_{\textrm{triplets}}&=
\begin{pmatrix}
m_{\widetilde{27}} & -2 \RED{c_1} \kappa _3 & -\frac{2 \sqrt{10} \lambda _1^{2/3} \lambda _2^{1/3} \left(2 m_{27} m_{351'} \kappa _3 \kappa_4-m_{\widetilde{27}}^2 \lambda _3 \lambda_4\right)}{\sqrt{3}m_{351'} \kappa _2 \lambda _3 \lambda _4} \\
 -\frac{m_{27} m_{351'} \kappa_4}{\RED{c_1} \lambda _3 \lambda_4} & m_{\widetilde{27}} & 0 \\
 \frac{\sqrt{2} m_{351'} \kappa _2}{3\sqrt{15} \lambda _1^{2/3} \lambda _2^{1/3}} & 0 & m_{\widetilde{27}}\\
\end{pmatrix}.
\end{align}
These matrices can also be found in equations~\eqref{equation:dt-mass-matrix} and \eqref{equation:triplet-mass-tilde}, but we have now plugged in the solution \eqref{equation:specific-first}-\eqref{equation:specific-last} and the DT fine-tuning from \eqref{equation:kappa1-fine-tuning}. A reminder: the VEV $c_1$ is determined by the polynomial~\eqref{equation:c1-polynomial}.

The terms with the $C$-coefficients come from the three Yukawa terms $\yd^{ij}$, $\yt^{ij}$ and $\ydt^{ij}$ in equation~\eqref{equation:W-full}. The barred $C$ coefficients come in pairs, e.g.~$\overline{C}_1$ and $\overline{C}_1'$, since the light state $\hat{L}$ is a linear combination of $L$ and $L'$, and similarly
$\hat{d}^c$ is a combination of $d^c$ and $d'^c$.

The coefficients $C$ are computed to be
\begin{align}
2\;C_1^{ijA}&=-\yd^{ij}\,\delta^{A}{}_{1}-\ydt^{ij}\,\delta^{A}{}_{1+12}+\tfrac{1}{2\sqrt{10}}\,\yt^{ij}\,\delta^{A}{}_{5}
-\tfrac{1}{2\sqrt{6}}\,\yt^{ij}\,\delta^{A}{}_{7}-\tfrac{1}{2\sqrt{3}}\,\yt^{ij}\,\delta^{A}{}_{12},\\
2\;C_2^{ijA}&=-\yd^{ij}\,\delta^{A}{}_{1}-\ydt^{ij}\,\delta^{A}{}_{1+12}+\tfrac{1}{2\sqrt{10}}\,\yt^{ij}\,\delta^{A}{}_{5}
-\tfrac{1}{2\sqrt{6}}\,\yt^{ij}\,\delta^{A}{}_{7}+\tfrac{2}{2\sqrt{3}}\,\yt^{ij}\,\delta^{A}{}_{12},\\
2\;\overline{C}_1^{ijA}&=-\yd^{ij}\,\delta^{A}{}_{2}-\ydt^{ij}\,\delta^{A}{}_{2+12}+
\tfrac{1}{2\sqrt{10}}\,\yt^{ij}\,\delta^{A}{}_{4}+\tfrac{1}{2\sqrt{2}}\,\yt^{ij}\,\delta^{A}{}_{8},\\
2\,\overline{C}_1'^{ijA}&=\phantom{-}\yd^{ij}\,\delta^{A}{}_{3}+\ydt^{ij}\,\delta^{A}{}_{3+12}-
\tfrac{1}{2\sqrt{10}}\,\yt^{ij}\,\delta^{A}{}_{9}-\tfrac{1}{2\sqrt{2}}\,\yt^{ij}\,\delta^{A}{}_{11},\\
2\;\overline{C}_2^{ijA}&=-\yd^{ij}\,\delta^{A}{}_{2}-\ydt^{ij}\,\delta^{A}{}_{2+12}+
\tfrac{1}{2\sqrt{10}}\,\yt^{ij}\,\delta^{A}{}_{4}-\tfrac{1}{2\sqrt{2}}\,\yt^{ij}\,\delta^{A}{}_{8},\\
2\,\overline{C}_2'^{ijA}&=\phantom{-}\yd^{ij}\,\delta^{A}{}_{3}+\ydt^{ij}\,\delta^{A}{}_{3+12}-
\tfrac{1}{2\sqrt{10}}\,\yt^{ij}\,\delta^{A}{}_{9}+\tfrac{1}{2\sqrt{2}}\,\yt^{ij}\,\delta^{A}{}_{11}.
\end{align}

Notice the different coefficients in front of $\delta^A{}_{12}$, a consequence of different Clebsch-Gordan coefficients
in Table \ref{table:pn-doublets-triplets}.

Integrating out the triplets $T_A$ and antitriplets $\overline{T}_A$ from the relevant terms, we obtain (to lowest order in the operators)
\begin{align}
\label{Wprotonfinal}
W&=-\Big(\overline{C}_1^{ijA}\,Q_i\,L_j+ \overline{C}_1'^{ijA} \, Q_i\,L'_j+\overline{C}_2^{ijA}\,d^c_i\,u^c_j+\overline{C}_2'^{ijA}\,d'^c_i\,u^c_j\Big)\big(\hat{\mathcal{M}}_{T}^{-1}\big)_{AB}
\Big(C_1^{klB}\, Q_k\,Q_l + C_2^{klB}\, u^c_k\,e^c_l\Big).
\end{align}

Note that we have written the inverse matrix $\hat{\mathcal{M}}_{T}^{-1}$ with a hat. A triplet mode and an antitriplet mode are massless, which means the block $\mathcal{M}_{\textrm{triplets}}$ cannot be inverted. The massless modes correspond to the would-be Goldstone bosons in the Higgs mechanism; these unphysical degrees of freedom can be rotated out of the Yukawa terms by a gauge transformation, which is equivalent to plugging a zero for their field value. This is formally equivalent to introducing a mass term for these modes, integrating them out, and then pushing the introduced mass to infinity, so they decouple from the theory. A basis independent ansatz for the computation of the inverse of the physical degrees of freedom, while automatically decoupling the would-be Goldstone bosons, is
\begin{align}
\hat{\mathcal{M}}_{T}^{-1}=\lim_{M\rightarrow\infty}\left(\mathcal{M}^{AB}+M\,f^A e^B\right)^{-1},
\end{align}
where $e^A$ and $f^A$ are the components of right and left null eigenvectors of $\mathcal{M}$, respectively. They need not be normalized, since the normalization factors can be absorbed into $M$. In our basis, we can take
\begin{align}
e^A&=
\frac{3 \sqrt{2} \RED{c_1} \lambda_1^{2/3} \lambda _2^{1/3}}{m_{351'}}\,\delta^{A}{}_{3}+\frac{\lambda _1^{1/3}}{\sqrt{6} \lambda _2^{1/3}}\,\delta^A{}_{4}+
\frac{6 \RED{c_1}^2 \lambda _1^{2/3} \lambda _2^{1/3} \lambda _3}{m_{351'}^2}\,\delta^A{}_{6}+\frac{\sqrt{5} \lambda _1^{1/3}}{\sqrt{6}\lambda _2^{1/3}}\,\delta^A{}_{8}+\delta^A{}_{10},\\
f^A&=\frac{3 m_{27} \lambda _1^{1/3} \lambda _2^{2/3}}{\sqrt{2} \RED{c_1} \lambda _3 \lambda _4}\,\delta^{A}{}_{3}+
\frac{\lambda _2^{1/3}}{\sqrt{6} \lambda _1^{1/3}}\,\delta^A{}_{4}+\frac{3 m_{27}^2 \lambda_1^{1/3} \lambda _2^{2/3}}{2 \RED{c_1}^2 \lambda _3^2 \lambda _4}\,\delta^A{}_{6}+\frac{\sqrt{5} \lambda _2^{1/3}}{\sqrt{6}\lambda _1^{1/3}}\,\delta^A{}_{8}+\delta^A{}_{10}.
\end{align}

Although formally elegant, this method is hard to implement since it requires to first explicitly invert a large matrix and only then
take the limit $M\to\infty$. An equivalent but more explicit procedure would be to rotate the $(N+1)$-dimensional system of
triplets into the $N$-dimensional part orthogonal to the Nambu-Goldstone zero mode. Let the normalized right and left Nambu-Goldstone eigenstates respectively be
\begin{align}
\frac{e}{|e|}&\equiv
\bem
\sqrt{1-\alpha^\dagger\alpha} \cr
\alpha
\eem,&
\frac{f}{|f|}&\equiv
\bem
\sqrt{1-\bar{\alpha}^\dagger\bar{\alpha}} \cr
\bar{\alpha}
\eem,
\end{align}

\noi
with the columns $\alpha=\alpha^a$ and $\bar{\alpha}=\bar{\alpha}^a$, $a=1,\ldots,N$.

The unitary $(N+1)\times(N+1)$ matrix

\beq
\label{Ualpha}
U(\alpha)=
\bem
\sqrt{1-\alpha^\dagger\alpha} & -\alpha^\dagger \cr
\alpha & 1-\frac{\alpha\alpha^\dagger}{1+\sqrt{1-\alpha^\dagger\alpha}}
\eem
\eeq

\noi
then transforms the old basis
\beq
T_A\to U_A{}^B(\alpha)T_B,
\eeq

\noi
where now $T_B=(T_0,T_a)$ with $T_0$ the would-be Nambu-Goldstone triplet. The $\overline{T}_A$ are analogously transformed by $U(\bar{\alpha})$. The choice of $U$ represents just one simple possibility of choosing the transformations matrix; it is not unique since we could have composed it with an arbitrary rotation in the orthogonal complement of the zero mode (space of $T_a$'s). Dropping the zero modes $T_0, \overline{T}_0$, equation~\eqref{Wproton} can now be written as

\begin{align}
W\big|_{\textrm{proton}}&=T_a\,(U^T)^a{}_A(\alpha)\,(\mathcal{M}_T)^{AB}\,U_B{}^b(\bar\alpha)\,\overline{T}_b
+ T_a\,(U^T)^a{}_A(\alpha)\left(C_1^{ijA}\; Q_i\,Q_j + C_2^{ijA}\; u^c_i\,e^c_j\right) \nonumber\\
&\quad+\left( \overline{C}_1^{ijB}\;Q_i\,L_j+ \overline{C}_1'^{ijB} Q_i\,L'_j+\overline{C}_2^{ijB}\;d^c_i\,u^c_j+\overline{C}_2'^{ijB}\;d'^c_i\,u^c_j\right)\,U_B{}^b(\bar\alpha)\,\overline{T}_b.
\end{align}

Defining now a $N\times N$ invertible matrix

\beq
(m_T)^{ab}\equiv (U^T)^a{}_A\,(\alpha)({\cal M}_T)^{AB}\,U_B{}^b(\bar\alpha),
\eeq

\noi
we arrive to (\ref{Wprotonfinal}) with the inverse of the $\mathcal{M}_{\textrm{triplets}}$ block given by

\beq
\big(\hat{\mathcal{M}}_{T}^{-1}\big)_{AB}=U_A{}^a(\bar\alpha)\,\left(m_T^{-1}\right)_{ab}\,(U^T)^b{}_B(\alpha).
\eeq

Finally, we have to project onto the light matter superfields. Remember that
the $27_F$'s contain a vector-like pair of quarks and leptons in the $10$ of $\SO(10)$. This $10$ mixes with the $16$, so
$d^c$ mixes with $d'^c$ and $L$ mixes with $L'$. This is rotated to the basis of light and heavy states with the help of the matrix $\mathcal{U}$ in equation~\eqref{equation:U-matrix}. Generic particles $q$ and $q'$ can be decomposed into light and heavy states $q_l$ and $q_H$, respectively, with
\begin{align}
\begin{pmatrix}
q& q'\\
\end{pmatrix}&=
\begin{pmatrix}
q_l& q_H\\
\end{pmatrix}\; \mathcal{U}.
\end{align}
This implies the following projections to the light states $\hat{d}^c$ and $\hat{L}$:
\begin{align}
d^c_i&=\big[(1+\tfrac{4}{9}X_0^\ast\,X_0^T)^{-1/2}\big]_{i}{}^{j}\,\hat{d}^c_j+\ldots,\\
d'^c_i&=\big[\tfrac{2}{3}X_0^T\;(1+\tfrac{4}{9}X_0^\ast\,X_0^T)^{-1/2}\big]_{i}{}^{j}\,\hat{d}^c_j+\ldots,\\
L_i&=\big[(1+X_0^\ast\,X_0^T)^{-1/2}\big]_i{}^{j}\,\hat{L}_j+\ldots,\\
L'_i&=\big[-X_0^T\;(1+X_0^\ast\,X_0^T)^{-1/2}\big]_{i}{}^{j}\,\hat{L}_j+\ldots,
\end{align}
where $X_0$ is defined in equation~\eqref{equation:X0-definition}.

Writing in terms of only light states (those at the scale $m_W$) and only for the lepton and baryon number violating operators, we get the following low-energy effective operators for $D=5$ proton decay:
\begin{align}
W\big|_{\textrm{proton}}&=-\Big[\big(\overline{C}_1^{inA}-\overline{C}_1'^{imA}(X_0^T)_{m}{}^{n}\big)\big[(1+X_0^\ast X_0^T)^{-1/2}\big]_{n}{}^{j}\;(\hat{\mathcal{M}}_T^{-1})_{AB}\; C_1^{klB}\Big]\;Q_i \hat{L}_j Q_k Q_l\nonumber\\
&\quad -\Big[\big(\overline{C}_2^{njA}+\tfrac{2}{3}\overline{C}_2'^{mjA}(X_0^T)_{m}{}^{n}\big)\big[(1+\tfrac{4}{9}X_0^\ast X_0^T)^{-1/2}\big]_{n}{}^{i}\;(\hat{\mathcal{M}}_T^{-1})_{AB}\; C_2^{klB}\Big]\;\hat{d}^c_i u^c_j u^c_k e^c_l.
\end{align}

In spite of the fact that the final expression is rather complicated, we can draw some general conclusions and leave
the numerical analysis in combination with the study of the Yukawa part for a future publication.
\begin{itemize}
\item
Since we are
in $E_6$ with several possible heavy thresholds, there is no necessary light color triplet as in the minimal
renormalizable $\SU(5)$ with low-scale supersymmetry \cite{Murayama:2001ur}.
\item
Only some elements of the inverse matrix need to be small, similar to the $\SO(10)$ case, which also has multiple triplet contributions. Notice that the triplets in $351'$, $\overline{27}$ and $\overline{\widetilde{27}}$ do not couple to the matter fields, and neither do some triplets in $\overline{351'}$ (seen from the $C$-coefficients). 
\item
The final expressions are functions of a number of parameters: the masses, the $\lambda$ and $\kappa$ parameters, as well as three Yukawa matrices. Since the constraints on these parameters come from the fit to a smaller number of parameters of the SM Yukawas, there will likely be some residual freedom in parameter space, which would allow for proton decay supression.
\end{itemize}
\noi
All these reasons make nucleon decay amplitude
suppressions probable. Finally, if all this fails, we can still use some version of a (moderately)
split supersymmetric spectrum.

\section{Conclusions\label{section:conclusions}}

This paper represents a first attempt to write down and solve a realistic $E_6$ model. We concentrated
on a supersymmetric and renormalizable case and found strong evidence that such a model
includes the fields $351'+\overline{351'}+27+\overline{27}$ to spontaneously break  $E_6$ into the SM,
another pair of $27+\overline{27}$ for the MSSM Higgs fields, and three copies of matter $27$'s.
To simplify the analysis we made two assumptions: an extra $\mathbb{Z}_2$ symmetry which automatically
preserves R-parity, and some vanishing couplings of the superfields that contain the MSSM Higgses.
Although the first assumption is probably unavoidable, the second may not be needed.

We noticed some interesting features which are not usually encountered in theories with lower groups:

\begin{itemize}

\item
The existence of asymmetric solutions: although the D-terms are satisfied by the natural solution $|\phi_i|=|\bar\phi_i|$
with $i$ going over all complex Higgs representations, the F-terms are not, unless the same VEVs vanish or there are
some fine-tuned relations among the superpotential parameter. We avoided such an assumption, and found
an asymmetric solution $|\phi_i|\ne|\bar\phi_i|$.

\item
The minimal sector that breaks into the SM could not describe the MSSM Higgses in spite
of the fact that it contains fields with the right quantum numbers. The reason is the impossibility of
performing a realistic DT splitting.

\item
The automatic presence of $3$ vector-like families, which makes the analysis of the Yukawa sector nonlinear.

\end{itemize}

An obvious problem with this type of models is the Landau pole which occurs at a scale $\Lambda$
less than one order of magnitude above the GUT scale (the gauge beta function is -153 compared to -109
of the minimal SO(10) \cite{Aulakh:1982sw,Clark:1982ai,Aulakh:2003kg}). So even if we believe that for some reason
gravity will not produce higher dimensional operators suppressed by inverse powers of $M_{Planck}$,
a consistent perturbative treatment of our type of models should assume the absence of operators
suppressed by inverse powers of $\Lambda$ as well. There is nothing we can say in defense of this,
except that well studied SO(10) models have similar problems.

There are several open questions, which are beyond the scope of this paper.
First, it would be interesting to check what happens
if other terms in the light Higgs superpotential are introduced, i.e. if the terms linear or cubic
in "tilde" fields appear. Second, since the neutrino masses need a slightly lower see-saw scale than
the GUT scale, the usual approximate one-step supersymmetric unification
\cite{Dimopoulos:1981yj,Ibanez:1981yh,Marciano:1981un,Einhorn:1981sx}
may be at risk due to large representations involved, and the knowledge of the
mass spectrum may turn out to be necessary. Third, more elegant solutions to the doublet-triplet splitting
problem can be looked for: both the missing VEV \cite{Dimopoulos:1981xm,Babu:1993we} and the missing
partner \cite{Masiero:1982fe,Babu:2006nf,Babu:2011tw} however seem to need the $650$: a full minimization
with this field should thus be performed. Fourth, a thorough study of the
$E_6$ could be studied at the non-renormalizable level: in this
case we could consider an asymptotically free theory with $78+27+\overline{27}$ only in the Higgs sector.
Fifth, non-supersymmetric theories could be considered, where the Higgs sectors are typically more
complicated, and intermediate states are mandatory. Finally, although we checked some simple cases with
$78$, there are still some possibilities for the Higgs sector to consider, for example $78+351+\overline{351}$
or $78+351'+\overline{351}'$. With the techniques described in this paper all these and other
issues can be attacked. We leave them for the future.

\section*{Acknowledgments}
Special thanks go to Debajyoti Choudhury and Goran Senjanovi\'c, with whom B.B. started to think about
$E_6$ quite some years ago. B.B. would like to thank CETUP* (Center for Theoretical Underground Physics
and Related Areas), supported by the US Department of Energy under Grant No. DE-SC0010137 and by the
US National Science Foundation under Grant No. PHY-1342611, for its hospitality and partial support during
the 2013 Summer Program, during which part of this work has been done. We are grateful
to Kaladi Babu, Ilia Gogoladze, St\'{e}phane Lavignac and Zurab Tavartkiladze for discussion
as well as correspondence and encouragement.
This work has been supported by the Slovenian Research Agency.

\appendix

\section{Detailed identification of various states\label{Appendix:particle-notation}}
It is possible to refer to states in various $E_6$ representations by using familiar labels of particles. The fundamental $27$ representation is $16+10+1$ in $\SO(10)$ language, so we denote the various SM irreducible representations in the following way:
\begin{itemize}
\item The $16$ contains $Q\sim (3,2,+\tfrac{1}{6})$, $L\sim (1,2,-\tfrac{1}{2})$, $u^c\sim (\overline{3},1,-\tfrac{2}{3})$, $d^c\sim (\overline{3},1,+\tfrac{1}{3})$, $e^c\sim (1,1,1)$ and $\nu^c\sim (1,1,0)$.
\item The $10$ contains $L'^c\sim (1,2,+\tfrac{1}{2})$, $L'\sim (1,2,-\tfrac{1}{2})$, $d'^c\sim (\overline{3},1,+\tfrac{1}{3})$ and $d'\sim (3,1,-\tfrac{1}{3})$. These are vector-like pairs of leptons and quarks.
\item The $\SO(10)$ singlet is labeled by $s\sim (1,1,0)$.
\end{itemize}

The conjugate representation of $27$ contains exactly the conjugate SM representations of the listed ones. We denote them by bars, so the particle content of $\overline{27}$ is labeled by $\bar{Q}$, $\bar{L}$, $\bar{d}^c$, $\bar{u}^c$, $\bar{e}^c$, $\bar{\nu}^c$, $\bar{L}'$, $\bar{L}'^c$, $\bar{d}'^c$, $\bar{d}'$ and $\bar{s}$.

Similarly, we can use this particle notation also for other irreducible representations. The representation $351'$ is contained in the symmetric product of $27\times 27$, so we label the states with two successive labels of the $27$, while suppressing  manifest symmetricity in our notation; to get the $351'$ in the $27\times 27$ matrix, we also have to project out the $\overline{27}$ with the $d$-tensor, but that does not change the rules of notation. Due to simplicity, we also suppress any color or weak indices from our notation, but summation over them is sometimes implicit.

As an example, consider the antitriplet $(\overline{3},1,+\tfrac{1}{3})$ contained in the product of two $(3,2,+\tfrac{1}{6})$: we label it simply by $QQ$, but this written out explicitly would be $\varepsilon_{abc}Q^{bi}Q^{cj}\varepsilon_{ij}$, where $a,b,c=1,2,3$ are color indices, $i,j=1,2$ are weak indices and $\varepsilon$ are the antisymmetric tensors; note that the only remaining free index is the lower $a$, which indicates the state is a $\overline{3}$ under $\SU(3)_C$. Another example is the antidoublet $(1,2,-\tfrac{1}{2})$ in $Qu^c$: explicitly, we actually mean $Q^{ai} (u^c)_a+(u^c)_a Q^{ai}$ (note also the symmetricity), with the same convention for the color and weak indices as before. Notice that we do not write (overall) normalization factors in particle notation; we use this notation only to identify the relevant states (which is non-trivial in the case of $351'$, since it is necessary to project out the $\overline{27}$), while in explicit computations we always use properly normalized states, such that for the doublet and triplets (see Table~\ref{table:pn-doublets-triplets}), we have
\begin{align}
351'^\ast{}_{\mu\nu}\;351'^{\mu\nu}+\overline{351'}^\ast{}^{\mu\nu}\;\overline{351'}_{\mu\nu}&= \sum_{i=1}^{11}\big(|D_i|^2+|\overline{D}_i|^2 \big) +\sum_{j=1}^{12}\big(|T_j|^2+|\overline{T}_j|^2\big) +\ldots.
\end{align}

We list the labels and particle identifications for the relevant fields in various tables below. The SM singlets are listed in Table~\ref{table:pn-singlets}, while the weak doublets and triplets relevant for DT splitting are listed in Table~\ref{table:pn-doublets-triplets}. We write them out explicitly in particle notation only in the unbarred representations, since the corresponding states in the conjugate representation would have the same form, but with ordinary letters substituted by barred letters. Also, the list of weak triplets contributing to type II seesaw has already been presented in Table~\ref{table:pn-weak-triplets}.

Note that the states are part of distinct representations, so they need to be orthogonal. This can be easily checked by the reader from particle notation (i.e.~the different two-label states are orthogonal basis vectors), but one should not forget about our convention of suppression of indices; what looks like a single term at first glance may indeed be a sum of multiple terms.

{\small
\begin{table}[h!]
\caption{Singlet labels and their identification in particle notation.\label{table:pn-singlets}}
\vskip 0.2cm
\begin{tabular}{lll@{\hspace{1cm}}lll}
\toprule
label&{\footnotesize $E_6\supseteq\SO(10)\supseteq\SU(5)$}&p.n.&label&{\footnotesize $E_6\supseteq\SO(10)\supseteq\SU(5)$}&p.n.\\\midrule
$c_1$&$\phantom{0}27\phantom{'}\supseteq\phantom{00}1\supseteq\phantom{0}1$&$s$&$d_1$&$\phantom{0}\overline{27}\phantom{'}\supseteq\phantom{00}1\supseteq\phantom{0}1$&$\bar{s}$\\
$c_2$&$\phantom{0}27\phantom{'}\supseteq \phantom{0}16\supseteq \phantom{0}1$&$\nu^c$&$d_2$&$\phantom{0}\overline{27}\phantom{'}\supseteq \phantom{0}\overline{16}\supseteq \phantom{0}1$&$\bar{\nu}^c$\\\addlinespace
$e_1$&$351'\supseteq 126\supseteq\phantom{0}1$&$\nu^c\nu^c$&$f_1$&$\overline{351'}\supseteq \overline{126}\supseteq\phantom{0}1$&$\bar{\nu}^c\bar{\nu}^c$\\
$e_2$&$351'\supseteq\phantom{0}16\supseteq\phantom{0}1$&$\nu^c s$&$f_2$&$\overline{351'}\supseteq\phantom{0}\overline{16}\supseteq\phantom{0}1$&$\bar{\nu}^c\bar{s}$\\
$e_3$&$351'\supseteq\phantom{00}1\supseteq\phantom{0}1$&$s\,s$&$f_3$&$\overline{351'}\supseteq\phantom{00}1\supseteq\phantom{0}1$&$\bar{s}\,\bar{s}$\\
$e_4$&$351'\supseteq\phantom{0}54\supseteq 24$&$L'L'^c-\tfrac{2}{3}d'^c d'$&$f_4$&$\overline{351'}\supseteq\phantom{0}54\supseteq 24$&$\bar{L}'\bar{L}'^c-\tfrac{2}{3}\bar{d}'^c \bar{d}'$\\
$e_5$&$351'\supseteq \overline{144}\supseteq 24$&$L\phantom{'}L'^c-\tfrac{2}{3}d^c\phantom{'} d'$&$f_5$&$\overline{351'}\supseteq 144\supseteq 24$&$\bar{L}\phantom{'}\bar{L}'^c-\tfrac{2}{3}\bar{d}^c\phantom{'} \bar{d}'$\\\bottomrule
\end{tabular}
\end{table}
}

{\small
\begin{table}[h!]
\caption{Doublet and triplet labels and identification in particle notation.\label{table:pn-doublets-triplets}}
\begin{longtable}{llll@{\hspace{0.5cm}}p{4cm}}
\toprule
label&{\footnotesize $E_6\supseteq\SO(10)\supseteq\SU(5)$}&label&{\footnotesize $E_6\supseteq\SO(10)\supseteq\SU(5)$}&doublet in p.n.\par triplet in p.n.\\\midrule
$D_1,T_1$&$\phantom{0}27\phantom{'}\supseteq\phantom{0}10\supseteq\phantom{0}5$&$\overline{D}_1,\overline{T}_1$&$\phantom{0}\overline{27}\phantom{'}\supseteq\phantom{0}10\supseteq\phantom{0}\overline{5}$&$L'^c$\par $d'$\\\addlinespace
$\overline{D}_2,\overline{T}_2$&$\phantom{0}27\phantom{'}\supseteq\phantom{0}10\supseteq\phantom{0}\overline{5}$&$D_2,T_2$&$\phantom{0}\overline{27}\phantom{'}\supseteq\phantom{0}10\supseteq\phantom{0}5$&$L'$\par $d'^c$\\\addlinespace
$\overline{D}_3,\overline{T}_3$&$\phantom{0}27\phantom{'}\supseteq\phantom{0}16\supseteq\phantom{0}\overline{5}$&$D_3,T_3$&$\phantom{0}\overline{27}\phantom{'}\supseteq\phantom{0}\overline{16}\supseteq\phantom{0}5$&$L$\par $d^c$\\\addlinespace
$D_4,T_4$&$351'\supseteq\phantom{0}10\supseteq\phantom{0}5$&$\overline{D}_4,\overline{T}_4$&$\overline{351'}\supseteq\phantom{0}10\supseteq\phantom{0}\overline{5}$&$Q d^c-L e^c-4L'^c \nu^c$\par $QL-u^c d^c-4d's$\\\addlinespace
$\overline{D}_5,\overline{T}_5$&$351'\supseteq\phantom{0}10\supseteq\phantom{0}\overline{5}$&$D_5,T_5$&$\overline{351'}\supseteq\phantom{0}10\supseteq\phantom{0} 5$&$Q u^c-L\nu^c-4L' s$\par $u^c e^c-d^c\nu^c+QQ-4d'^c s$\\\addlinespace
$\overline{D}_6,\overline{T}_6$&$351'\supseteq\phantom{0}16\supseteq\phantom{0}\overline{5}$&$D_6,T_6$&$\overline{351'}\supseteq\phantom{0}\overline{16}\supseteq\phantom{0}5$&$-L s$\par $-d^c s$\\\addlinespace
$\overline{D}_7,\overline{T}_7$&$351'\supseteq 126\supseteq\phantom{0}\overline{5}$&$D_7,T_7$&$\overline{351'}\supseteq \overline{126}\supseteq\phantom{0} 5$&$-Qu^c-3L\nu^c$\par $-u^c e^c-3 d^c\nu^c-QQ$\\\addlinespace
$D_8,T_8$&$351'\supseteq126\supseteq 45$&$\overline{D}_8,\overline{T}_8$&$\overline{351'}\supseteq\overline{126}\supseteq \overline{45}$&$Qd^c+3Le^c$\par $QL+u^c d^c$\\\addlinespace
$D_9,T_9$&$351'\supseteq\overline{144}\supseteq\phantom{0}5$&$\overline{D}_9,\overline{T}_9$&$\overline{351'}\supseteq 144\supseteq\phantom{0}\overline{5}$&$-Qd'^c+4L'^c\nu^c+L'e^c$\par $-QL'+u^c d'^c+4d' \nu^c$\\\addlinespace
$\overline{D}_{10},\overline{T}_{10}$&$351'\supseteq\overline{144}\supseteq\phantom{0}\overline{5}$&$D_{10},T_{10}$&$\overline{351'}\supseteq 144\supseteq\phantom{0} 5$&$-L'\nu^c$\par $-d'^c\nu^c$\\\addlinespace
$D_{11},T_{11}$&$351'\supseteq\overline{144}\supseteq 45$&$\overline{D}_{11},\overline{T}_{11}$&$\overline{351'}\supseteq 144\supseteq\overline{45}$&$-dd'^c-3e'e^c$\par $-QL'-u^c d'^c$\\\addlinespace
$\phantom{D_{1}}\overline{T}_{12}$&$351'\supseteq 126\supseteq \overline{50}$&$\phantom{D_{1}}T_{12}$&$\overline{351'}\supseteq \overline{126}\supseteq 50$&$/$\par $2u^c e^c- QQ$\\\bottomrule
\end{longtable}
\end{table}
}

\newpage
\section{Details of the vacuum\label{Appendix:vacuum}}
We present here some details of the SM (supersymmetric) vacuum, obtained by the particular solution in equations~\eqref{equation:specific-first}-\eqref{equation:specific-last}.
\subsection{Gauge boson masses}
The masses of the gauge bosons $A_\mu{}^{a}$ can be computed explicitly. The gauge boson mass terms can be written as
\begin{align}
\mathcal{L}_{mass}&=g^2 A_\mu^{\;a} M^{ab} A^{\mu\,b},
\end{align}
where $g$ is the $E_6$ gauge coupling constant, and the the mass-square matrix $M^{ab}$ is defined as
\begin{align}
M^{ab}&\equiv(\hat{t}^{a} 27)^\dagger (\hat{t}^{b} 27)+(\hat{t}^{a} \overline{27})^\dagger (\hat{t}^{b} \overline{27})+(\hat{t}^{a} 351')^\dagger (\hat{t}^{b} 351')+(\hat{t}^{a} \overline{351'})^\dagger (\hat{t}^{b} \overline{351'}).
\end{align}
The symbol $\hat{t}^a$ denotes the action of the $a$-th generator on the representation. Knowing the explicit form of the generators $t^a$ in the fundamental representation, and using $E_6$ tensor methods, we can compute the matrix $M^{ab}$ explicitly, and determine the squares of the gauge boson masses by diagonalization. The results for the specific solution with $c_2=d_2=e_2=f_2=e_4=f_4=0$ are given in Table~\ref{table:gauge-boson-masses}. The nonzero VEVs of the solution are not plugged in.

Notice that the $\SU(5)$ singlets in $45$ and $1$ of $\SO(10)$ mix among themselves, as do the singlets in the $16$ and $\overline{16}$ of $\SO(10)$. The $\SU(5)$ $10$'s in the $45$ and $16$ do not mix, and neither do the $\overline{10}$'s in the $45$ and $\overline{16}$. Plugging $f_5=e_5=0$, only the $\SU(5)$ singlets are nonzero, so we get $24$ massless gauge bosons, while the other bosons in $\SU(5)$ representations get the same mass. The same thing happens if only the $\SO(10)$ singlets $c_1$, $d_1$, $e_3$ and $f_3$ are nonzero: we get $45$ massless gauge bosons, and the others' masses get grouped according to $\SO(10)$ representations.

\newpage
\def\THICK{0.2pt}
\begin{table}[h!]
\caption{Masses-squared of gauge bosons in SM representations using the ansatz \hbox{$c_2=d_2=e_2=f_2=e_4=f_4=0$.}\label{table:gauge-boson-masses}}
\vskip 0.2cm
\centering
\scalebox{0.98}{
\begin{tabular}{p{1.5cm}p{1.5cm}p{2cm}p{8cm}}
\toprule
$\SO(10)\supset$&$\SU(5)\supset$&$\textrm{SM}\supset$&$\textrm{(mass)}^2/g^2$\\\midrule
$45$&$24$&$(8,1,\,0)$&$0$\\\addlinespace
$45$&$24$&$(1,3,\,0)$&$0$\\\addlinespace
$45$&$24$&$(1,1,\,0)$&$0$\\\midrule[\THICK]
$45$&$24$&$(3,2,+\tfrac{5}{6})$&$\tfrac{5}{6}|\RED{e_5}|^2+\tfrac{5}{6}|\RED{f_5}|^2$\\
&&$(\overline{3},2,-\tfrac{5}{6})$&\\\midrule[\THICK]
$45$&$10$&$(3,2,+\tfrac{1}{6})$&$|\RED{e_1}|^2+|\RED{f_1}|^2+\tfrac{1}{2}|\RED{e_5}|^2+\tfrac{1}{2}|\RED{f_5}|^2$\\
&$\overline{10}$&$(\overline{3},2,-\tfrac{1}{6})$&\\\midrule[\THICK]
$45$&$10$&$(\overline{3},1,-\tfrac{2}{3})$&$|\RED{e_1}|^2+|\RED{f_1}|^2+\tfrac{1}{2}|\RED{e_5}|^2+\tfrac{1}{2}|\RED{f_5}|^2$\\
&$\overline{10}$&$(3,1,+\tfrac{2}{3})$&\\\midrule[\THICK]
$45$&$10$&$(1,1,+1)$&$|\RED{e_1}|^2+|\RED{f_1}|^2+\tfrac{1}{2}|\RED{e_5}|^2+\tfrac{1}{2}|\RED{f_5}|^2$\\
&$\overline{10}$&$(1,1,-1)$&\\\midrule[\THICK]
$16$&$10$&$(3,2,+\tfrac{1}{6})$&$\tfrac{1}{2}|\RED{c_1}|^2+\tfrac{1}{2}|\RED{d_1}|^2+|\RED{e_3}|^2+|\RED{f_3}|^2+\tfrac{5}{6}|\RED{e_5}|^2+\tfrac{5}{6}|\RED{f_5}|^2$\\
$\overline{16}$&$\overline{10}$&$(\overline{3},2,-\tfrac{1}{6})$&\\\midrule[\THICK]
$16$&$10$&$(\overline{3},1,-\tfrac{2}{3})$&$\tfrac{1}{2}|\RED{c_1}|^2+\tfrac{1}{2}|\RED{d_1}|^2+|\RED{e_3}|^2+|\RED{f_3}|^2$\\
$\overline{16}$&$\overline{10}$&$(3,1,+\tfrac{2}{3})$&\\\midrule[\THICK]
$16$&$10$&$(1,1,+1)$&$\tfrac{1}{2}|\RED{c_1}|^2+\tfrac{1}{2}|\RED{c_1}|^2+|\RED{e_3}|^2+|\RED{f_3}|^2$\\
$\overline{16}$&$\overline{10}$&$(1,1,-1)$&\\\midrule[\THICK]
$16$&$\phantom{0}\overline{5}$&$(\overline{3},1,+\tfrac{1}{3})$&$\tfrac{1}{2}|\RED{c_1}|^2+\tfrac{1}{2}|\RED{d_1}|^2+|\RED{e_1}|^2+
|\RED{f_1}|^2+$\\
$\overline{16}$&$\phantom{0}5$&$(3,1,-\tfrac{1}{3})$&$\qquad\qquad +|\RED{e_3}|^2+|\RED{f_3}|^2+\tfrac{1}{2}|\RED{e_5}|^2+\tfrac{1}{2}|\RED{f_5}|^2$\\\midrule[\THICK]
$16$&$\phantom{0}\overline{5}$&$(1,2,-\tfrac{1}{2})$&$\tfrac{1}{2}|\RED{c_1}|^2+\tfrac{1}{2}|\RED{d_1}|^2+|\RED{e_1}|^2+
|\RED{f_1}|^2+$\\
$\overline{16}$&$\phantom{0}5$&$(1,2,+\tfrac{1}{2})$&$\qquad\qquad +|\RED{e_3}|^2+|\RED{f_3}|^2+\tfrac{1}{2}|\RED{e_5}|^2+\tfrac{1}{2}|\RED{f_5}|^2$\\\midrule[\THICK]
$45$\par $\phantom{0}1$&$\phantom{0}1$\par $\phantom{0}1$&$(1,1,\,0)$\par $(1,1,\,0)$&
They mix:\vskip -0.6cm
\begin{equation*}
\tfrac{2}{3}\Big((A+B)\pm \sqrt{(A+B)^2-\tfrac{15}{4}AB}\;\Big),
\end{equation*}
\vskip -0.3cm
$A\equiv 4 |\RED{e_1}|^2+4|\RED{f_1}|^2+|\RED{e_5}|^2+|\RED{f_5}|^2$\par $B\equiv 4|\RED{e_3}|^2+4|\RED{f_3}|^2+|\RED{c_1}|^2+|\RED{d_1}|^2$
\\\midrule[\THICK]
$16$\par $\overline{16}$&$\phantom{0}1$\par $\phantom{0}1$&$(1,1,\,0)$\par $(1,1,\,0)$&
They mix:\vskip -0.6cm
\begin{equation*}
\tfrac{1}{2}\Big((C+D)\pm \sqrt{(C-D)^2+16|E|^2}\;\Big),
\end{equation*}
\vskip -0.3cm
$C\equiv |\RED{c_1}|^2+ 2 |\RED{f_1}|^2 + 2 |\RED{e_3}|^2+ |\RED{e_5}|^2$\par
$D\equiv |\RED{d_1}|^2+ 2 |\RED{e_1}|^2 + 2 |\RED{f_3}|^2 +|\RED{f_5}|^2$\par
$E\equiv \RED{e_1} \RED{e_3}^\ast + \RED{f_1}^\ast \RED{f_3}$
\\\bottomrule
\end{tabular}
}
\end{table}

\newpage
\subsection{No flat directions check}
In order to check that our solution is an isolated point and that there are no flat directions in the $F$-terms, we check the mass matrix of the SM VEV-acquiring singlets in our model. The relevant singlets live in the representations $27$, $\overline{27}$, $351'$ and $\overline{351'}$ of the Higgs sector. We label the singlets by $s_x$, where $x$ is the label of the VEV; our singlets are therefore $s_{c_i}$, $s_{d_i}$, $s_{e_j}$ and $s_{f_j}$, where $i=1,2$ and $j=1,\ldots,5$.

The mass term can be written as
\begin{align}
\frac{1}{2}\,
\begin{pmatrix}
s_{d_i}&s_{c_i}&s_{f_j}&s_{e_j}\\
\end{pmatrix}
\;\mathcal{M}_{\textrm{singlets}}\;
\begin{pmatrix}
s_{c_i}\\
s_{d_i}\\
s_{e_j}\\
s_{f_j}\\
\end{pmatrix},
\end{align}
\noindent
where $\mathcal{M}_{\textrm{singlets}}$ is the matrix
\begin{align}
\hspace{-0.5cm}
\left(
\begin{smallmatrix}
m_{27} & 0 & 2 \lambda_4 \RED{e_3} & \sqrt{2} \lambda_4 \RED{e_2} & 0 & \sqrt{2} \lambda_4 \RED{d_2} & 2 \lambda_4 \RED{d_1} & 0 & 0 & 0 & 0 & 0 & 0 & 0 \\
 0 & m_{27} & \sqrt{2} \lambda_4 \RED{e_2} & 2 \lambda_4 \RED{e_1} & 2 \lambda_4 \RED{d_2} & \sqrt{2} \lambda_4 \RED{d_1} & 0 & 0 & 0 & 0 & 0 & 0 & 0 & 0 \\
 2 \lambda_3 \RED{f_3} & \sqrt{2} \lambda_3 \RED{f_2} & m_{27} & 0 & 0 & 0 & 0 & 0 & 0 & 0 & \sqrt{2} \lambda_3 \RED{c_2} & 2 \lambda_3 \RED{c_1} & 0 & 0 \\
 \sqrt{2} \lambda_3 \RED{f_2} & 2 \lambda_3 \RED{f_1} & 0 & m_{27} & 0 & 0 & 0 & 0 & 0 & 2 \lambda_3 \RED{c_2} & \sqrt{2} \lambda_3 \RED{c_1} & 0 & 0 & 0 \\
 0 & 2 \lambda_3 \RED{c_2} & 0 & 0 & m_{351'} & 0 & 0 & 0 & 0 & 0 & 0 & 0 & 0 & 6 \lambda_2 \RED{f_5} \\
 \sqrt{2} \lambda_3 \RED{c_2} & \sqrt{2} \lambda_3 \RED{c_1} & 0 & 0 & 0 & m_{351'} & 0 & 0 & 0 & 0 & 0 & 0 & -3 \sqrt{2} \lambda_2 \RED{f_5} & -3 \sqrt{2} \lambda_2 \RED{f_4} \\
 2 \lambda_3 \RED{c_1} & 0 & 0 & 0 & 0 & 0 & m_{351'} & 0 & 0 & 0 & 0 & 0 & 6 \lambda_2 \RED{f_4} & 0 \\
 0 & 0 & 0 & 0 & 0 & 0 & 0 & m_{351'} & 0 & 0 & -3 \sqrt{2} \lambda_2 \RED{f_5} & 6 \lambda_2 \RED{f_4} & 6 \lambda_2 \RED{f_3} & -3 \sqrt{2} \lambda_2 \RED{f_2} \\
 0 & 0 & 0 & 0 & 0 & 0 & 0 & 0 & m_{351'} & 6 \lambda_2 \RED{f_5} & -3 \sqrt{2} \lambda_2 \RED{f_4} & 0 & -3 \sqrt{2} \lambda_2 \RED{f_2} & 6 \lambda_2 \RED{f_1} \\
 0 & 0 & 0 & 2 \lambda_4 \RED{d_2} & 0 & 0 & 0 & 0 & 6 \lambda_1 \RED{e_5} & m_{351'} & 0 & 0 & 0 & 0 \\
 0 & 0 & \sqrt{2} \lambda_4 \RED{d_2} & \sqrt{2} \lambda_4 \RED{d_1} & 0 & 0 & 0 & -3 \sqrt{2} \lambda_1 \RED{e_5} & -3 \sqrt{2} \lambda_1 \RED{e_4} & 0 & m_{351'} & 0 & 0 & 0 \\
 0 & 0 & 2 \lambda_4 \RED{d_1} & 0 & 0 & 0 & 0 & 6 \lambda_1 \RED{e_4} & 0 & 0 & 0 & m_{351'} & 0 & 0 \\
 0 & 0 & 0 & 0 & 0 & -3 \sqrt{2} \lambda_1 \RED{e_5} & 6 \lambda_1 \RED{e_4} & 6 \lambda_1 \RED{e_3} & -3 \sqrt{2} \lambda_1 \RED{e_2} & 0 & 0 & 0 & m_{351'} & 0 \\
 0 & 0 & 0 & 0 & 6 \lambda_1 \RED{e_5} & -3 \sqrt{2} \lambda_1 \RED{e_4} & 0 & -3 \sqrt{2} \lambda_1 \RED{e_2} & 6 \lambda_1 \RED{e_1} & 0 & 0 & 0 & 0 & m_{351'}
\end{smallmatrix}
\right)
\end{align}
Plugging in the specific solution from equations~\eqref{equation:specific-first}-\eqref{equation:specific-last}, the matrix gets $4$ massless modes. Since the $E_6\rightarrow\textrm{SM}$ breaking also breaks $4$ of the $5$ SM singlet generators in the $78$, we expect that any solution of the equations of motion breaking to the SM will automatically produce $4$ massless singlet modes of scalars due to the Higgs mechanism. Any additional massless modes would correspond to a flat direction of the superpotential around the specific vacuum. Since there are no additional massless singlet modes, there are no flat directions in our solution.

\section{The doublet-triplet splitting\label{Original-DT-splitting}}

Suppose we have a Higgs sector consisting of fields $351'+\overline{351'}+27+\overline{27}$ and we want the SM Higgs to live in both the $27$ and $\overline{351'}$, similar to how the Higgs lives both in the $10$ and $\overline{126}$ in the $\SO(10)$ model. The mass terms connecting doublets $(1,2,+\tfrac{1}{2})$ to antidoublets $(1,2,-\tfrac{1}{2})$ and the triplets $(3,1,-\tfrac{1}{3})$ to antitriplets $(\overline{3},1,+\tfrac{1}{3})$ will come from the breaking part of the superpotential in equation~\eqref{equation-superpotential}. The mass matrices of the doublets and triplets in the fermionic $27_F^i$ are distinct and do not mix with the mass matrices in the breaking sector due to $\mathbb{Z}_2$ matter parity.

The list of all the doublets and triplets, along with our label conventions, is already compiled in Table~\ref{table:pn-doublets-triplets}. There are $11$ doublet/antidoublet pairs and $12$ triplet/antitriplet pairs in the breaking sector, so the doublet and triplet mass matrices are $11\times 11$ and $12\times 12$, respectively.

Writing the doublet and triplet mass terms as

\begin{align}
\begin{pmatrix}D_1&\cdots& D_{11}\\\end{pmatrix}
 \mathcal{M}_{\textrm{doublets}}
\begin{pmatrix}\overline{D}_1\\\vdots \\\overline{D}_{11}\\\end{pmatrix}+
\begin{pmatrix}T_1&\cdots& T_{12}\\\end{pmatrix}
 \mathcal{M}_{\textrm{triplets}}
\begin{pmatrix}\overline{T}_1\\\vdots \\\overline{T}_{12}\\\end{pmatrix},
\end{align}
the mass matrices $\mathcal{M}_{\textrm{doublets}}$ and $\mathcal{M}_{\textrm{triplets}}$ can be compactly written as

\begin{align}
\scalebox{0.8}{
$
\hspace{-0.5cm}
\left(
\begin{smallmatrix}
m_{27} & \alpha\lambda_{3}\frac{ \RED{f_4} }{\sqrt{15}}-6 \lambda_{5}\RED{c_1} & \alpha\lambda_{3}\frac{\RED{f_5}}{\sqrt{15}}+6 \lambda_{5} \RED{c_2} & -\sqrt{\frac{8}{5}} \lambda_{3} \RED{c_1} & 0 & 0 & 0 &0 & \sqrt{\frac{8}{5}} \lambda_{3} \RED{c_2} & 0 & 0 & 0 \\
\alpha\lambda_{4} \frac{\RED{e_4}}{\sqrt{15}}-6 \lambda_{6} \RED{d_1} &m_{27} & 0 & 0 & -\sqrt{\frac{8}{5}} \lambda_{4} \RED{d_1} & 0 & 0& 0 & 0 & -\sqrt{2} \lambda_{4} \RED{d_2} & 0 & 0 \\
\alpha\lambda_{4} \frac{\RED{e_5}}{\sqrt{15}}+6 \lambda_{6} \RED{d_2}& 0& m_{27} & 0 & -\lambda_{4}\frac{\RED{d_2}}{\sqrt{10}} & -\sqrt{2}\lambda_{4} \RED{d_1} & -\sqrt{\frac{3}{2}} \lambda_{4} \RED{d_2} & 0 & 0& 0 & 0 & 0 \\
-\sqrt{\frac{8}{5}} \lambda_{4} \RED{d_1} & 0 & 0 & m_{351'} &\alpha\sqrt{\frac{3}{5}} \lambda_{1} \RED{e_4} & 0 & 0 & 0 & 0 &-\alpha \frac{\sqrt{3}}{2}  \lambda_{1} \RED{e_5}& 0 & 0 \\
0 & -\sqrt{\frac{8}{5}} \lambda_{3} \RED{c_1} & -\lambda_{3}\frac{\RED{c_2}}{\sqrt{10}} & \alpha\sqrt{\frac{3}{5}} \lambda_{2} \RED{f_4}& m_{351'} & 0 & 0 & 0 & -\alpha\frac{1}{4} \sqrt{\frac{3}{5}} \lambda_{2}\RED{f_5} & 0 & -\beta\frac{5\sqrt{3}}{4} \lambda_{2} \RED{f_5} & 0 \\
0 & 0 & -\sqrt{2} \lambda_{3} \RED{c_1} & 0 & 0 & m_{351'} & 0 & 0 &\alpha\frac{\sqrt{3}}{2} \lambda_{2} \RED{f_4}& 0 & \beta\frac{\sqrt{15}}{2} \lambda_{2} \RED{f_4}& 0 \\
0 & 0 & -\sqrt{\frac{3}{2}} \lambda_{3} \RED{c_2} & 0 & 0 & 0 & m_{351'}& \beta\sqrt{5} \lambda_{2} \RED{f_4} & -\alpha\frac{3}{4} \lambda_{2}\RED{f_5} & 0 & \beta\frac{\sqrt{5}}{4} \lambda_{2} \RED{f_5}& 0 \\
0 & 0 & 0 & 0 & 0 & 0 & \beta\sqrt{5} \lambda_{1} \RED{e_4} & m_{351'}& 0 & -\beta\frac{\sqrt{15}}{2}\lambda_{1} \RED{e_5} & 0 &\alpha\sqrt{10} \lambda_{1} \RED{e_4} \\
\sqrt{\frac{8}{5}} \lambda_{4} \RED{d_2} & 0 & 0 & 0 & -\alpha\frac{1}{4}\sqrt{\frac{3}{5}} \lambda_{1} \RED{e_5}  & \alpha\frac{\sqrt{3}}{2}\lambda_{1} \RED{e_4}& -\alpha\frac{3}{4} \lambda_{1} \RED{e_5}& 0 & m_{351'} & 0 & 0 & 0 \\
0 & -\sqrt{2} \lambda_{3} \RED{c_2} & 0 & -\alpha\frac{\sqrt{3}}{2} \lambda_{2}\RED{f_5} & 0 & 0 & 0 & -\beta\frac{\sqrt{15}}{2} \lambda_{2}\RED{f_5}& 0 & m_{351'} & 0 & 0 \\
0 & 0 & 0 & 0 & -\beta \frac{5\sqrt{3}}{4} \lambda_{1} \RED{e_5} &\beta \frac{\sqrt{15}}{2} \lambda_{1} \RED{e_4} & \beta\frac{\sqrt{5}}{4} \lambda_{1} \RED{e_5}& 0 & 0 & 0 & m_{351'} &\alpha\sqrt{10} \lambda_{1} \RED{e_5}  \\
0 & 0 & 0 & 0 & 0 & 0 & 0 & \alpha\sqrt{10} \lambda_{2} \RED{f_4} & 0 &0 & \alpha\sqrt{10} \lambda_{2} \RED{f_5} & m_{351'}\\
\end{smallmatrix}
\right) .\label{equation:dt-mass-matrix}
$
}
\end{align}

For the triplet matrix take $\alpha=\beta=2$, while for the doublet matrix remove the last row and column and take $\alpha=-3$ and $\beta=-\sqrt{3}$.

Note the form of the above matrix $\mathcal{M}_{ij}$: for $i,j=1,2,3$, the doublets and triplets come from the pair $27+\overline{27}$, while indices $i,j=4,\ldots,n$ refer to fields coming from the pair $351'+\overline{351'}$, where $n=11$ and $n=12$ for doublets and triplets, respectively. That means the matrix has the block form

\begin{align}
\begin{pmatrix}
M_{3\times 3}&M_{3\times (n-3)}\\
M_{(n-3)\times 3}&M_{(n-3)\times (n-3)}\\
\end{pmatrix},
\end{align}
\noindent
where the blocks are in addition to the mass terms populated by the following invariants:
\begin{itemize}
\item Block $M_{3\times 3}$ is populated by $27^2\times\langle 27,\overline{351'}\rangle$ and $\overline{27}^2\times\langle\overline{27},351'\rangle$.
\item Blocks $M_{3\times (n-3)}$ and $M_{(n-3)\times 3}$ are populated by $27\times \overline{351'}\times\langle 27\rangle$ and $\overline{27}\times351'\times\langle\overline{27}\rangle$.
\item Block $M_{(n-3)\times (n-3)}$ is populated by $351'^2\times\langle 351'\rangle$ and $\overline{351'}^2\times\langle \overline{351'}\rangle$.
\end{itemize}
Notice that $c_i$ and $d_i$ are $\SU(5)$ singlets, so matrix entries with these VEVs for the doublets and triplets are the same, since both come from the same $\SU(5)$ representation. The $\SU(5)$ singlets $e_1,e_2,e_3$ and $f_1,f_2,f_3$ from $351'$ and $\overline{351'}$ do not come into play, but the VEVs  $e_4,e_5,f_4,f_5$ do. The latter VEVs are $\langle 24\rangle$ under $\SU(5)$, so the matrix entries containing these VEVs differentiate between the doublets and triplets, as one would expect, so values of $\alpha$ and $\beta$ between doublets and triplets differ. The value for the $\alpha$ comes directly from the VEV $\langle 24\rangle\propto\mathrm{diag}(2,2,2,-3,-3)$ which couples $5$'s to $\overline{5}$'s, while $\beta$ has the unusual value $-\sqrt{3}$ for triplets. This is due to the fact that the $\beta$ entries couple the doublets/triplets in the $45$ and $\overline{5}$ (or equivalently $\overline{45}$ and $5$) of $\SU(5)$; the $-\sqrt{3}$ is due to the normalization of the doublets and triplets in the representations $45$ and $\overline{45}$.

Performing doublet-triplet splitting would involve a fine-tuning of parameters $m_{351'}$, $m_{27}$ and $\lambda_i$, such that we get a massless doublet mode, while keeping all the triplets heavy. Solving the equations of motion and plugging their solution into the mass matrices (i.e. using equations~\eqref{general-first}-\eqref{general-last}), we automatically get a massless doublet/antidoublet and triplet/antitriplet mode. This should come as no surprise, since the breaking $E_6\rightarrow \textrm{SM}$ involves breaking the generators transforming as $16+\overline{16}$ of $\SO(10)$, which contain exactly one doublet/antidoublet pair and one triplet/antitriplet pair: the doublet and triplet massless modes of scalars are eaten up by the corresponding gauge bosons due to the Higgs mechanism. We thus have the extra complication of already having a massless doublet and triplet mode, so DT splitting involves making a second doublet mode massless.

In principle, the massless modes can be straightforwardly extracted from the scalar squared-mass matrices \hbox{$\mathcal{M}_{\textrm{doublets}}^\dagger\mathcal{M}_{\textrm{doublets}}$} and \hbox{$\mathcal{M}_{\textrm{triplets}}^\dagger \mathcal{M}_{\textrm{triplets}}$.} In our case, however, squaring a matrix would unnecessarily complicate the calculation, so we use methods which work on the matrices $\mathcal{M}_{\textrm{doublets}}$ and $\mathcal{M}_{\textrm{triplets}}$ themselves.

Already having a massless mode in $\mathcal{M}^\dagger\mathcal{M}$ implies
\begin{align}
\det\mathcal{M}&=0.
\end{align}
The condition for another massless mode in $\mathcal{M}^\dagger\mathcal{M}$ can be written as
\begin{align}
\mathrm{Cond}(\mathcal{M})&:=\frac{\lim_{\epsilon\rightarrow 0}\det(\mathcal{M}+\epsilon I)/\epsilon}{\langle f|e \rangle}=0,
\end{align}
where $I$ is the identity matrix and $|e\rangle$ and $| f\rangle$ are the already present right and left zero-mass eigenmodes of $\mathcal{M}$:
\begin{align}
\mathcal{M}|e\rangle=\mathcal{M}^\dagger|f\rangle&=0.
\end{align}

Using our vacuum solution and confirming
\begin{align}
\det \mathcal{M}_{\textrm{doublets}}=\det\mathcal{M}_{\textrm{triplets}}&=0,
\end{align}

\noindent
the DT splitting conditions read
\begin{align}
\mathrm{Cond}(\mathcal{M}_{\textrm{doublets}})&=\frac{1}{72}\;m_{351'}^9 m_{27}\;\frac{\lambda_3 \lambda_4}{\lambda_1 \lambda_2}=0,\\
\mathrm{Cond}(\mathcal{M}_{\textrm{triplets}})&=\frac{4}{243}\;m_{351'}^{10} m_{27}\;\frac{\lambda_3 \lambda_4}{\lambda_1 \lambda_2}\neq 0.
\end{align}

We see that due to the simplicity of the conditions, which are just a product of the Lagrangian parameters, we cannot perform a fine-tuning on the doublets independently from the triplets: an extra massless doublet necessarily implies an extra massless triplet. The usual procedure of DT splitting via fine-tuning is therefore not possible in this case. We cure this problem of the model by adding an extra $\widetilde{27}+\overline{\widetilde{27}}$ pair.

Although explicit calculation shows DT splitting is not possible without the tilde fields, we are unable to find a clear-cut reason, which would explain --- without calculation --- why the usual method of fine-tuning fails. Note that the mass matrices by themselves do not have this feature: the impossibility of DT
splitting shows itself only after solving the $F$-term equations of motion and plugging in the solutions. In the following, we list peculiar details and possible reasons, a combination of which might contribute to this inability:
\begin{itemize}
\item It seems DT splitting is a problem only for solutions breaking to the SM. It is possible to fine-tune in the alternative solution $\langle 27\rangle=\langle\overline{27}\rangle=0$, which breaks to Pati-Salam.
\item Without the tilde fields, the SM Higgs would live in representations already involved in the $E_6$ breaking: these representations would thus acquire both GUT and EW scale VEVs.
\item We already have a massless doublet and triplet mode present in the mass matrices due to the Higgs mechanism.
\item There are only a few specific places in the mass matrices (in the $M_{3\times 3}$ block), where there is a sum of the type $A\,\langle 1\rangle+B\,\langle 24\rangle$, which enables DT splitting in the simplest $\SU(5)$ case. When plugging in a specific solution with the ansatz $c_2=d_2=e_4=f_4=0$, one of the two terms always disappears. Also, the mass matrices contain a lot of zero entries.
\end{itemize}

As a final point, we allude to the missing partner mechanism (MP mechanism) \cite{Masiero:1982fe,Babu:2006nf,Babu:2011tw}
for DT splitting as it pertains to our model. At first glance, the MP mechanism would seem promising for our model, since we already have one more triplet than a doublet due to the presence of $50$ and $\overline{50}$ of $\SU(5)$ in the representations $351'$ and $\overline{351'}$. The above explicit computation already shows that the mechanism is not at work in our case, which is perhaps expected: the MP mechanism involves a specific setup of fields and form of the mass matrix, so it would be too optimistic to expect it for free in a general setup. To implement the MP mechanism, it is necessary to have the $50$ of $\SU(5)$ with the triplet but not the doublet, but also the $75$, which couples the triplet in the $50$ with the antitriplet in the $\overline{5}$. The lowest dimensional $E_6$ representation, which contains the $75$ of $\SU(5)$, is the $650$. In our case, simplicity therefore seems to dictate to forego the MP mechanism, and just add the extra $\widetilde{27}+\overline{\widetilde{27}}$ pair.

\end{document}